\documentclass[aps,prb]{revtex4-2}
\usepackage[utf8]{inputenc}
\DeclareUnicodeCharacter{2010}{-}
\DeclareUnicodeCharacter{2013}{--}
\DeclareUnicodeCharacter{2014}{---}
\DeclareUnicodeCharacter{2019}{'{} }

\usepackage{amsmath,amssymb}
\usepackage{graphicx}
\usepackage{bm}
\usepackage{xcolor}
\usepackage{eurosym}
\usepackage{amsmath,amssymb}
\usepackage{graphicx}
\usepackage{dcolumn}
\usepackage{bm}
\usepackage{color}
\usepackage{multirow}
\usepackage{float}
\usepackage{array}
\usepackage{dcolumn}
\usepackage{booktabs}
\usepackage[detect-all]{siunitx}
\usepackage{makecell}
\usepackage{xcolor}
\usepackage{yfonts}
\usepackage{makecell}
\begin{document}

\title{Bound Trions in Two-Dimensional Monolayers: A Review}

\author{Roman~Ya.~Kezerashvili}
\affiliation{New York City College of Technology, The City University of New York, Brooklyn, NY, USA}
\affiliation{The Graduate School and University Center, The City University of New York, New York, NY, USA}
\affiliation{Long Island University, Brooklyn, NY, USA}

\date{\today}

\begin{abstract}

Trions—Coulomb-bound three-particle excitations composed of two like-charge carriers and one oppositely charged carrier—are central quasiparticles in two-dimensional semiconductors. Reduced dielectric screening and quantum confinement strongly enhance their binding energies, making them robust and experimentally accessible. This review surveys theoretical and experimental advances in trion physics, emphasizing rigorous few-body approaches and the role of dielectric environment, anisotropy, and external electric and magnetic fields. We analyze computational methods for describing trions in two-dimensional configuration spaces and discuss how reduced dimensionality modifies their structure and stability. Connections to many-body phenomena, including screening, Landau-level mixing, and exciton–polaron crossover, are also highlighted.

\end{abstract}

\maketitle

\begin{center}
{CONTENTS}
\end{center}

\vspace{0.5em}

\noindent
I. \ INTRODUCTION\\
II.\ OVERVIEW OF TRIONS BEFORE THE GRAPHENE ERA\\
III.\ FROM BULK AND QUANTUM-WELL SEMICONDUCTORS TO ATOMICALLY THIN MATERIALS\\
IV.\ CHARGED-CARRIER INTERACTION IN 2D MONOLAYERS\\
V.\ EXPERIMENTAL OBSERVATION OF TRIONS IN 2D MONOLAYERS\\
\hspace*{1.5em}A.\ Trions in TMDC: Experimental trion binding energies\\
\hspace*{1.5em}B.\ Trions in anisotropic monolayers\\
\hspace*{1.5em}C.\ Trions in Xenes\\
VI.\ THEORETICAL APPROACHES AND FORMALISM\\
\hspace*{1.5em}A.\ Deterministic few-body methods\\
\hspace*{3.0em}1.\ Variational method\\
\hspace*{3.0em}2.\ Stochastic variational method\\
\hspace*{3.0em}3.\ Exact (direct) diagonalization method\\
\hspace*{3.0em}4.\ Gaussian expansion method\\
\hspace*{1.5em}B.\ Diffusion quantum Monte Carlo\\
\hspace*{1.5em}C.\ Path-integral Monte Carlo\\
\hspace*{1.5em}D.\ Faddeev equations approach for trions\\
\hspace*{1.5em}E.\ Hyperspherical harmonics formalism\\
VII.\ BENCHMARKING THEORETICAL AND COMPUTATIONAL APPROACHES TO TRIONS IN 2D MONOLAYERS\\
VIII.\ TRIONS IN EXTERNAL ELECTRIC AND MAGNETIC FIELDS\\
\hspace*{1.5em}A.\ Trions in 2D materials under external electric field\\
\hspace*{1.5em}B.\ Trions in 2D materials under external magnetic field\\
IX.\ CONCLUDING REMARKS


\section{Introduction}
\label{Sect_1}

The conceptual roots of trions lie in the theory of excitons, introduced by Frenkel in 1931 and later refined by Wannier and Mott, who established the framework for Wannier--Mott excitons in semiconductors \cite{Wannier1937,Mott1968}. An exciton is a bound state of an electron and a hole held together by the Coulomb attraction between their opposite charges. In typical semiconductors, the exciton Bohr radius—associated with the relative motion of the electron and hole—is much larger than the lattice constant. Consequently, excitons in inorganic semiconductors are well described as weakly bound Wannier--Mott excitons \cite{Bassani1975,LaRocca2003}, in contrast to tightly bound Frenkel excitons characteristic of molecular and organic crystals \cite{Knoester2003}.

The concept of charged excitonic complexes, now commonly referred to as trions, predates the modern era of atomically thin materials by several decades. It emerged naturally from early developments in exciton physics and from the study of Coulomb-bound few-body systems in semiconductors. Once excitons were recognized as elementary quasiparticles in insulating and semiconducting crystals, it became clear that an exciton could, in principle, bind an additional electron or hole, forming a stable three-particle complex.

The first systematic theoretical description of such complexes was provided by Lampert \cite{Lampert1958} in 1958. He introduced charged excitons as effective-mass analogues of the negative hydrogen ion in solids and demonstrated that an exciton can bind an extra electron or hole, giving rise to negatively charged ($X^-$) or positively charged ($X^+$) excitons. These three-particle bound states were later termed \textit{trions} to emphasize their composite nature. Early theoretical studies treated trions as few-body Coulomb systems within the effective-mass approximation, identifying the carrier effective masses and dielectric screening as key parameters governing their stability and binding energies.

This review presents a comprehensive overview of the few-body physics of trions in reduced-dimensional semiconductors, with particular emphasis on two-dimensional (2D) materials. It is organized as follows.

Sections~\ref{Sect_2} and~\ref{Sect_3} provide a historical and conceptual overview of trion physics, beginning with the pre-graphene era in bulk and quantum-well semiconductors and progressing toward atomically thin systems.

Section~\ref{Sect_4} discusses charge--charge interactions in 2D monolayers, with emphasis on dielectric screening and reduced-dimensional effects. A systematic summary of experimental observations of trions in 2D monolayers is presented in Section~\ref{Sect_5}.

Section~\ref{Sect_6} introduces the general three-body theoretical formalism within the effective-mass framework and reviews major computational and analytical approaches, including variational and stochastic variational methods, exact diagonalization, Gaussian expansion techniques, diffusion and path-integral quantum Monte Carlo methods, Faddeev equations, and the hyperspherical harmonics method.

Section~\ref{Sect_7} benchmarks theoretical approaches against reported trion binding energies in 2D monolayers. Section~\ref{Sect_8} examines the influence of external electric and magnetic fields on trions in 2D materials, highlights the distinctive role of group-IV monolayers, and discusses magnetotrion dynamics, emphasizing the necessity of exact Landau-level mixing and the intrinsic coupling between center-of-mass and internal degrees of freedom for reliable modeling in magnetic fields. Concluding remarks are presented in Section~\ref{Sect_9}.

\section{Overview of Trions Before the Graphene Era}
\label{Sect_2}

Lampert’s original prediction of charged excitons \cite{Lampert1958} represents the canonical starting point of virtually all modern discussions of trion physics. During the 1960s and 1970s, subsequent theoretical studies examined excitons bound to impurities and free carriers in bulk semiconductors. In particular, Moskalenko \cite{Moskalenko1960}, building on the exciton framework summarized by Mott \cite{Mott1964}, presented one of the earliest microscopic treatments of impurity-bound excitons, analyzing their Coulomb binding to charged centers and their radiative recombination. It was demonstrated that such complexes give rise to distinct luminescence lines shifted relative to free-exciton emission. Although this work focused on impurity-bound states rather than free three-particle complexes, it clearly established that excitons can bind additional charges while remaining optically active, thereby anticipating the concept of charged excitons and providing an early foundation for modern trion theory.

In the late 1970s, extremely weakly bound three-particle Coulomb complexes were observed in bulk semiconductors such as Ge and Si \cite{Taniguchi1975,Narita1976PRL,Taniguchi1976SSC}. These findings motivated quantitative theoretical descriptions of negative donor ion ($D^-$) states in bulk materials. In particular, Kamimura \textit{et al.}~\cite{Kamimura1979} and Chang \cite{Chang1982} developed effective-mass-based theories incorporating realistic multivalley band-structure effects for Ge and Si, treating $D^-$ states as weakly bound three-body Coulomb systems. Together, these works established benchmark descriptions of bulk three-particle binding that informed and guided later studies of trions in reduced-dimensional semiconductor systems.

A rigorous three-body treatment of charged excitonic complexes in bulk semiconductors was presented in Ref.~\cite{Filikhin2018} using the Faddeev-equation framework, focusing on the stability of negatively and positively charged trions under realistic effective-mass and dielectric-screening conditions. The study shows that while negatively charged trions, $X^-$, may exist as extremely weakly bound states under favorable circumstances, positively charged trions, $X^+$, are intrinsically unstable in bulk materials, even when realistic mass ratios and Coulomb interactions are taken into account. This fundamental asymmetry originates from strong dielectric screening, the absence of confinement, and the large effective masses of holes in bulk semiconductors: a positively charged trion, composed of two holes and one electron, exhibits strongly suppressed hole delocalization due to the heavy hole masses, leading to a more compact spatial distribution of the two holes and enhanced direct Coulomb repulsion in the positively charged trion. Therefore, the hole--hole Coulomb repulsion cannot be compensated by electron-mediated attraction. As a result, $X^+$ complexes are energetically unstable against dissociation, explaining their systematic absence in bulk-semiconductor experiments.

Thus, in three-dimensional materials, strong dielectric screening significantly reduced the trion binding energies, typically to the scale of a few meV or less. As a consequence, experimental identification of trions in bulk systems remained elusive for many years.

Early bulk evidence for extremely weak three-body Coulomb binding in semiconductors (e.g., $D^-$ negative-ion analogs in doped Ge/Si) preceded the high-quality heterostructure era. The situation changed substantially in
the 1990s with the development of high-quality semiconductor heterostructures and quantum wells. Reduced dimensionality enhances Coulomb interactions and suppresses screening, thereby stabilizing few-body excitonic complexes. Throughout the 1980s, variational and numerical calculations within different theoretical frameworks predicted that trions in quasi-two-dimensional systems should possess experimentally resolvable binding energies \cite{Whittaker1997}. These predictions were spectacularly confirmed in the early 1990s by photoluminescence experiments in GaAs/AlGaAs quantum wells, which provided the first unambiguous observation of negatively charged excitons \cite{Kheng1993}. In the 1990s, high-quality semiconductor quantum wells allowed precise studies of trion states using optical measurements \cite{Kheng1993,Finkelstein1995,Shields1995,Buhmann1995}. Subsequent experiments identified both $X^-$ and $X^+$ states and clarified their spin structure, optical selection rules, and magnetic-field dependence \cite{Finkelstein1996}.

By the late 1990s and early 2000s, trions had become a well-established element of semiconductor many-body physics. An analytically guided variational/adiabatic framework was developed for indirect trions in double quantum wells \cite{Suris2003}, demonstrating that interwell separation introduces a critical distance beyond which the trion becomes unbound. This framework provides practical scaling trends and estimates for GaAs- and ZnSe-based heterostructures. The binding mechanism is physically transparent: the electron mediates an effective attraction between the two holes, which persists only up to a tunable layer separation. Advanced theoretical approaches, including the method of hyperspherical harmonics \cite{Ruan1999}, configuration interaction method \cite{Dacal2002CI}, exact diagonalization \cite{Wojs2007} method, and quantum Monte Carlo simulations \cite{Filinov2004PIMC}, were employed to compute trion binding energies and wave functions with high accuracy \cite{Riva2000}. Experimental studies have been extended to a variety of material systems, such as CdTe- and ZnSe-based quantum wells, and have explored regimes of strong magnetic fields, where trions coexist and interact with fractional quantum Hall states \cite{Esser2000}.

To summarize, negatively, $X^-$, and positively, $X^+$, charged trions confined in quantum wells, and subjected to external magnetic or electric fields, have been the subject of extensive experimental \cite{Finkelstein1995,Shields1995,Buhmann1995,Warburton1997,Finkelstein1998,Ciulin2000,Es,Es1,Es2,Vanhoucke2002,A,BonitzCD,Sergeev2005,Astakhov2007} and theoretical \cite{Stebe1974,Stebe1977,Stebe1987,Stebe1989,Hawrylak1995,Stebe1997,Whittaker1997,Varga1999,Stebe2000,Riva2000,Stebe20001,Stebe20002,Dzyubenko2000,Varga2001,Wojs2001,Xie2001,Dacal2002,Varshni2001,Es1,Dacal20022,Szafran2005} investigations. While we cite representative works, these references do not exhaust the extensive literature on the subject.

Thus, before the graphene era, an extensive body of research had firmly established trions as genuine Coulomb-bound quasiparticles in quasi-two-dimensional systems, governed by universal few-body physics.

\section{From bulk and quantum-Well semiconductors to atomically thin materials
}
\label{Sect_3}
The historical development of trion physics in bulk and quantum-confined semiconductors laid the essential conceptual and methodological foundation for the modern resurgence of negatively and positively charged trions in atomically and molecularly thin materials. The discovery of graphene \cite{Novoselov2004} and, subsequently, monolayer transition-metal dichalcogenides (TMDCs)  \cite{Mak2010} and Xenes, monolayers of group-IV elements such as Si, Ge, Sn
revealed systems in which Coulomb interactions are dramatically enhanced due to extreme confinement, reduced dielectric screening, and tunable carrier densities. Xenes were first predicted theoretically \cite{Cahangirov2009,Xu2013} and subsequently realized experimentally on substrates, beginning with silicene \cite{Lalmi2010,Vogt2012} and followed by germanene \cite{Davila20142014} and stanene  \cite{Zhu2015}.

Despite graphene’s central role in launching two-dimensional materials research, it does not support robust optical trions under typical conditions, owing to its gapless Dirac spectrum and strong dielectric screening. Although theoretical studies have predicted possible excitonic or trionic instabilities in graphene under extreme conditions—such as very strong magnetic fields, substrate-induced band gaps, or substantially reduced screening—these regimes are not representative of standard experimental settings. Consequently, graphene is best viewed as a bridge material: although it does not host robust trions itself, it motivated the exploration of gapped two-dimensional semiconductors, notably TMDCs and Xenes, where reduced screening and finite band gaps enable the stabilization and direct observation of both negatively and positively charged trions. In contrast to conventional quantum wells, these atomically thin materials exhibit trion binding energies reaching tens of meV, rendering trions stable even at room temperature.
The dramatic resurgence of trion physics in TMDCs, Xenes, and other atomically thin monolayers after 2010 should therefore be understood not as a discovery \textit{ex nihilo}, but as a natural continuation of more than five decades of theoretical and experimental research on Coulomb-bound few-body complexes in semiconductors.

\section{Charged carrier interaction in 2D monolayers}
\label{Sect_4}

Since the experimental discovery of highly conductive graphene monolayers in
2004~\cite{Novoselov2004}, the field of condensed matter physics has experienced explosive growth in theoretical and experimental research in the realm of
2D materials. The isolation of graphene from bulk crystals of graphite marked the identification of the first member of a family of
2D layered materials, which have grown rapidly over the past ten years and
now includes insulators, semiconductors, semimetals, metals, and
superconductors \cite{Bhimanapati, Novoselov2}. 2D materials are commonly
defined as crystalline materials consisting of a single layer of atoms.
Most often, such materials are classified as either 2D allotropes of various
elements or compounds consisting of two or more ionically/covalently bonded
elements. These materials are crystalline solids with a high ratio between
their lateral size and thickness \cite{Velicky}. In van der Waals layered
materials, the atomic organization and bond strength in the 2D plane are
typically much stronger than in the third dimension (out-of-plane), where
they are bonded together by weak van der Waals interaction \cite{Novoselov2,
Jariwala}. Research on two-dimensional materials is primarily focused on graphene, transition-metal dichalcogenides, phosphorene, transition-metal trichalcogenides (TMTCs), and Xenes~\cite{Kormanyos,Joshua,Li1,Li2,Warren2015}. Among these, a particularly important and rapidly developing subclass is formed by buckled group-IV monolayers, collectively known as Xenes~\cite{Matthes2014,Molle2017,Brunetti2018}. This family includes silicene~\cite{Falko2012}, germanene~\cite{Davila2014RKkor}, stanene~\cite{Zhu2014RKkor}, and borophene~\cite{Mannix2015}, which are distinguished by their tunable band gaps under externally applied electric fields and offer additional flexibility for electronic and optoelectronic applications.

\begin{figure}
\centering
\begin{minipage}{0.58\linewidth}
\centering
\textbf{($a$)}\\[-2pt]
\includegraphics[width=5.7cm]{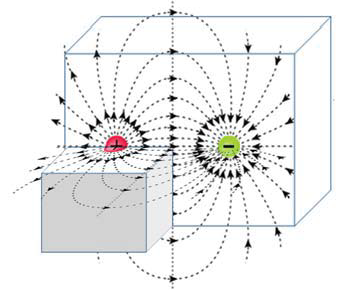}
\end{minipage}
\hfill
\begin{minipage}{0.38\linewidth}
\centering
\textbf{($b$)}\\[-2pt]
\includegraphics[width=4.0cm]{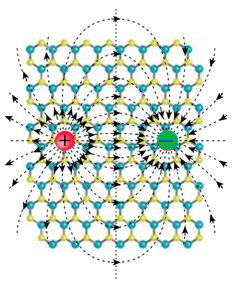}
\end{minipage}

\caption{Schematic representation of dipole electric fields in different dimensionalities. ($a$) The electric field of a dipole in a bulk 3D material. ($b$) Top-view schematic of a 2D monolayer, illustrating the in-plane electric field lines generated by an in-plane dipole. Fig. \ref{fig:2D_top}$a$ adapted from \cite{Kezerashvili2020FBProc.}.}
\label{fig:2D_top}
\end{figure}

The reduction of dimensionality has a profound and twofold effect on the
interaction of charged particles in monolayer materials. First, if the electron--hole interaction is described by the Coulomb potential in a
homogeneous dielectric environment and both charges are constrained to move in a plane, dimensional reduction affects the kinetic energy. In this case, the energy spectrum changes from the three-dimensional Rydberg series
$E_{3\mathrm{D}}\sim 2\varepsilon\hbar^{2}/n^{2}$ ($n=1,2,\ldots$) to the
two-dimensional form $E_{2\mathrm{D}}\sim
2\varepsilon\hbar^{2}/(n-1/2)^{2}$, resulting in a fourfold increase of the
ground-state binding energy and reflecting the suppression of kinetic energy
in two-dimensional excitons \cite{Kezerashvili2020FBProc.}.
Secondly, more importantly, dimensional reduction also modifies the potential energy of the electron--hole interaction. While the interaction remains electromagnetic, dielectric screening in two dimensions differs fundamentally from the three-dimensional case. In a monolayer, polarization
is confined to the plane, so that only the in-plane component of the induced electric field contributes to the interaction. As a result, the interaction remains a function of the in-plane separation $r$ but acquires a functional form that differs substantially from the Coulomb potential. Figure \ref{fig:2D_top} provides a schematic comparison between the electric field distribution of a dipole in a 3D bulk medium and the predominantly in-plane field lines characteristic of a dipole within a 2D monolayer.

The effective two-dimensional interaction between two point charges in a
layered dielectric environment was derived by Rytova~\cite{Rytova} and later
by Keldysh~\cite{Keldysh}. A modern re-derivation and quantitative validation of the Rytova--Keldysh
interaction, including its implications for excitonic states in atomically
thin materials, was later provided within a rigorous dielectric screening
framework~\cite{Cudazzo2011PRB}. This interaction, today known as the
Rytova--Keldysh (RK) potential, is given by
\begin{equation}
V(r_{eh})=-\frac{\pi ke^{2}}{\left(\varepsilon_{1}+\varepsilon_{2}\right)\rho_{0}}
\left[H_{0}\!\left(\frac{r}{\rho_{0}}\right)
-Y_{0}\!\left(\frac{r}{\rho_{0}}\right)\right].
\label{RKeldysh}
\end{equation}
Here $r$ is the in-plane electron--hole separation,
$k=1/(4\pi\epsilon_{0})$, $H_{0}$ and $Y_{0}$ are the Struve and Bessel
functions of the second kind, $\varepsilon_{1}$ and $\varepsilon_{2}$ are the
dielectric constants above and below the monolayer, and
$\rho_{0}=2\pi\zeta/[(\varepsilon_{1}+\varepsilon_{2})/2]$ is the screening
length determined by the two-dimensional polarizability $\zeta$. The RK
potential interpolates between a Coulomb tail
$V(r)\approx -ke^{2}/(\varepsilon r)$ for $r\gg\rho_{0}$ and a logarithmic
interaction
$V(r)\approx -\frac{ke^{2}}{\varepsilon\rho_{0}}
\left[\ln\!\left(\frac{r}{2\rho_{0}}\right)+\gamma\right]$ for
$r\ll\rho_{0}$, where $\gamma$ is the Euler constant. Thus, at short distances
the $1/r$ singularity is replaced by a much weaker logarithmic dependence.

Tuan et al.~\cite{Tuan2018PRB} critically examine the widely used Rytova--Keldysh potential for modeling Coulomb interactions in monolayer TMDCs. Although this potential has been extensively employed to describe excitons and trions in two-dimensional semiconductors, it does not consistently reproduce key experimental observations. Going beyond the conventional strictly 2D approximation, the authors ~\cite{Tuan2018PRB} develop a microscopic description of the Coulomb interaction that explicitly incorporates the finite thickness and intrinsic three-atomic-sheet structure of TMDC monolayers, consisting of a metal layer sandwiched between two chalcogen layers. The derived effective potential accounts for the nonuniform dielectric environment, the out-of-plane spatial separation of charges, and the orbital character of the band-edge states. They demonstrate that the Coulomb interaction deviates from the simple logarithmic-to-$1/r$ crossover of the RK model, particularly at short distances relevant to excitons and trions. The refined screening modifies binding energies and fine-structure splittings, providing a physically transparent framework that bridges atomistic and effective-mass descriptions for more accurate modeling in realistic dielectric environments.

The Ref. \cite{ExpZinkiewicz2022} focused on experimental observations of dielectric-environment effects on the magnetic-field-induced brightening of neutral and charged dark excitons in monolayer WSe$_2$, showing how encapsulation or substrate choice influences the optical visibility and activation of these spin-forbidden states.  The authors demonstrated that incorporating realistic dielectric screening and finite-thickness effects is crucial for accurately modeling Coulomb interactions and excitonic complexes in monolayer semiconductors, thereby achieving improved quantitative agreement between theoretical predictions and experimental observations.

Taken together, the results of Refs. \cite{Tuan2018PRB,ExpZinkiewicz2022} highlight the strong sensitivity of charge--charge interactions in two-dimensional materials to dielectric screening and confinement, thereby motivating the development of effective interaction potentials beyond the simple Coulomb form. In this context, modified Kratzer-type potentials, originally introduced in molecular physics by Kratzer~\cite{Kratzer1920}, have been proposed as analytically tractable models
for excitons in two-dimensional semiconductors~\cite{Molas2019PRL}. In
particular, this potential has been employed to model Rydberg excitons in dielectric environments, allowing the 2D Schrödinger
equation to be solved exactly and yielding closed-form expressions for the
eigenvalues and eigenfunctions~\cite{Molas2019PRL,Kezerashvili2023FBS,
Pedersen2023PRB,Kezerashvili2025IJTP}. Subsequent studies \cite{Pedersen2023PRB,Kezerashvili2025IJTP} demonstrated that the modified Kratzer potential admits exact solutions in two
dimensions and can be systematically applied to excitonic systems in layered
semiconductors.

The modified Kratzer potential takes the form~\cite{Molas2019PRL}
\begin{equation}
V(r) = -\frac{1}{4\pi\varepsilon_{0}}\frac{e^{2}}{\tilde{\rho_0}}
\left( \frac{r_{0}}{r} - g^{2}\frac{r_{0}^{2}}{r^{2}} \right),
\label{Kr25}
\end{equation}
where $\varepsilon_{0}$ is the vacuum permittivity,
$\tilde{\rho_0}=r_{0}\varepsilon$ is the effective screening length, and $\varepsilon$
is the dielectric constant of the surrounding medium. This potential
interpolates between different interaction regimes: for $r_{0}=1$ and $g=0$
it reduces to the Coulomb potential, while $g^{2}=1/2$ reproduces the original
Kratzer potential~\cite{Kratzer1920}.

The analytical solvability of the Kratzer-type potential~(\ref{Kr25}) makes it
a powerful tool for modeling electron--hole, electron--electron, and hole--hole interactions in low-dimensional
systems, in particular, to study $X^{-}$ and $X^+$ trions. A series of recent works has applied this approach to excitons and
related bound states in two-dimensional semiconductors, including TMDC
monolayers, demonstrating that it captures essential features of exciton
binding and dielectric screening while permitting exact or nearly exact
solutions of the Schrödinger equation~\cite{Kezerashvili2023FBS,
Kezerashvili2025IJTP}. Moreover, the modified Kratzer model has been used as a
basis for analytical estimates of exciton and trion binding energies and for
describing long-range interaction effects in layered materials, providing a
useful complement to more microscopic approaches~\cite{Molas2019PRL,
Pedersen2023PRB,Kezerashvili2025IJTP}.

\section{Experimental observation of trions in 2D monolayers}

\label{Sect_5}

\subsection{Trions in TMDC
}

The experimental investigation of trions in TMDC monolayers began shortly after the discovery
of strongly bound excitons in these atomically thin semiconductors.
The first unambiguous experimental measurements of trion binding energies
were reported in 2013 using low-temperature optical spectroscopy
\cite{Mak2013NatMat,Ross2013NatComm}.
In these early studies, photoluminescence (PL) spectra of the monolayer
MoS$_2$ and MoSe$_2$ revealed an additional emission peak at energies
below the exciton resonance.
This feature was identified as a charged exciton, formed by
binding an exciton with an extra electron or hole.

The trion binding energy is experimentally extracted from the energy
splitting between the neutral exciton and trion peaks,
$E_b = E(X^0) - E(X^\pm)$.
For monolayer MoS$_2$, the first measurements yielded
$E_b \simeq 18$--$20$~meV \cite{Mak2013NatMat}, already an order of magnitude
larger than typical trion binding energies in conventional semiconductor
quantum wells.
This immediately highlighted the crucial role of reduced dimensionality
and weak dielectric screening in enhancing Coulomb interactions in
2D materials.

\textbf{\textit{Experimental methods.}}
The dominant experimental techniques used to probe trions in TMDC monolayers
are optical in nature.
Low-temperature PL spectroscopy remains the most widely employed method,
owing to its sensitivity and relative simplicity.
Complementary information is obtained from reflectance contrast and
absorption spectroscopy, which directly probe excitonic resonances
\cite{Zhu2015SciRep,Jadczak2017NanoLett}.
In many experiments, electrostatic gating in field-effect transistor
geometries is used to tune the carrier density and control the charge state
of the excitonic complexes \cite{Ross2013NatComm,Chen2023PRB}.
This approach allows the selective formation of negative trions, $X^-$,
in electron-doped regimes and positive trions, $X^+$, in hole-doped regimes,
providing clear experimental confirmation of both species.

\textbf{\textit{Identification of $X^-$ and $X^+$ trions.}}
Gate-dependent PL measurements established that the trion peak evolves
systematically with doping polarity.
In monolayer MoSe$_2$ and WSe$_2$, both $X^-$ and $X^+$ trions were observed,
with binding energies typically in the range of 25--40~meV
\cite{Ross2013NatComm,Chen2023PRB}.
In many cases, the binding energies of positive and negative trions are
comparable, although small asymmetries are sometimes resolved and attributed
to differences in effective masses and band structure.
When charge discrimination is not experimentally possible, the reported
binding energy corresponds to the charged exciton peak without explicit
identification of its sign.

\textbf{\textit{Material dependence.}}
Systematic experimental studies revealed clear trends across the TMDC family.
Mo-based monolayers such as MoS$_2$ and MoSe$_2$ typically exhibit trion
binding energies in the range $\sim$18--30~meV
\cite{Mak2013NatMat,Ross2013NatComm}.
W-based materials, including WS$_2$ and WSe$_2$, generally show larger values,
often extending up to $\sim$40~meV
\cite{Zhu2015SciRep,Huang2016SciRep,Chen2023PRB}.
This enhancement is commonly attributed to stronger spin--orbit coupling
and heavier carrier effective masses.
In contrast, monolayer MoTe$_2$ exhibits somewhat smaller trion binding
energies, typically $\sim$14--27~meV, reflecting increased dielectric
screening and a reduced band gap
\cite{Yang2015ACSNano,Biswas2023arXiv}.

\textbf{\textit{Role of dielectric environment and disorder.}}
A key experimental finding is the strong sensitivity of trion binding
energies to the surrounding dielectric environment.
Comparisons between monolayers on SiO$_2$, encapsulated in hBN, or suspended
demonstrate that increased screening reduces the measured binding energy,
while encapsulation narrows spectral linewidths and improves sample quality
\cite{Jadczak2017NanoLett,ExpZinkiewicz2022}.
As a result, experimentally reported trion binding energies often span a
range rather than a single universal value for a given material.

\textbf{\textit{Temperature and carrier-density effects.}}
Trion resonances are robust at low temperatures and, in some materials, persist up to room temperature, albeit with significant broadening \cite{Mak2013NatMat,Hanbicki2016AO}. The first clear experimental observation of tightly bound trions in monolayer MoS$_2$ by Mak \textit{et al.}~\cite{Mak2013NatMat} demonstrated that the trion binding energy remains sufficiently large to survive thermal fluctuations at 300~K. Subsequent temperature-dependent photoluminescence studies have shown that while exciton and trion resonance energies redshift with increasing temperature due to electron--phonon interactions, the exciton--trion energy separation exhibits weak or moderate temperature dependence depending on the material system \cite{Ross2013NatComm,Christopher2017SciRep,Nguyen2024SciRep}. Increasing carrier density leads to enhanced screening and a gradual reduction of the exciton--trion energy splitting \cite{Mak2013NatMat,Efimkin2017}. At sufficiently high doping levels, the optical response crosses over to an exciton--polaron regime, reflecting the many-body nature of the electron--exciton interaction \cite{Sidler2017NatPhys}.

\textbf{\textit{Experimental significance.}}
The experimental discovery and systematic characterization of trions in
TMDC monolayers established these materials as a unique platform for
studying few-body and many-body Coulomb-correlated states in two dimensions.
The unusually large trion binding energies, combined with tunability via
gating, substrate engineering, and temperature, have made trions a central
element of the optical response of TMDCs and a benchmark for theoretical
three-body and polaronic descriptions of excitonic complexes in reduced
dimensionality.

\begin{figure}[b]
\begin{center}
\includegraphics[width=16cm]{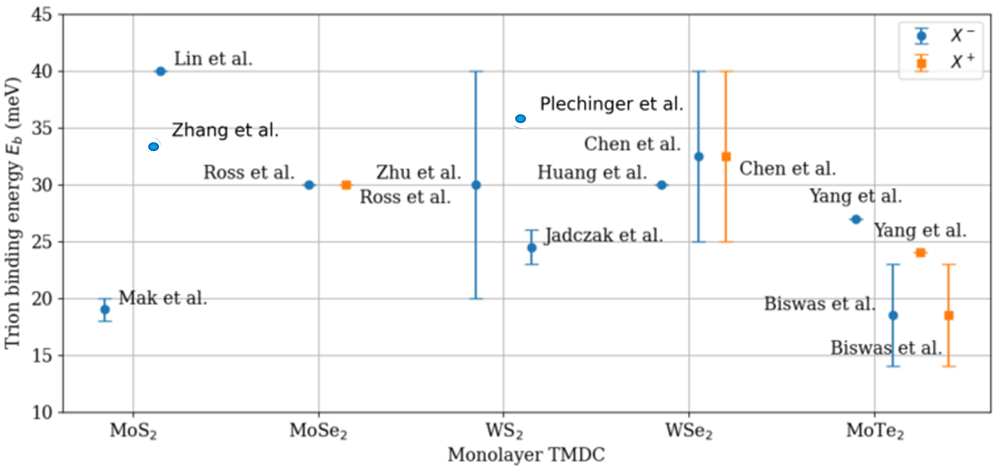}
\end{center}
\caption{Typical experimentally measured trion binding energies $E_b$ in
monolayer TMDCs from low-temperature optical spectroscopy. Points indicate
representative mean values; error bars show reported experimental ranges,
reflecting variations in dielectric environment, substrate, and carrier density.
Experimental trion binding energies in monolayer TMDCs are taken from Refs.
\cite{Mak2013NatMat,Ross2013NatComm,Zhu2015SciRep,Jadczak2017NanoLett,Chen2023PRB,Huang2016SciRep,Biswas2023arXiv,Yang2015ACSNano,Lin2014NanoLett,Plechinger2016,ZhangMS2}.
}
\label{TMDCtrion}
\end{figure}

Different experimental
groups \cite{ExpZinkiewicz2022,Mak2013NatMat,Ross2013NatComm,Lin2014NanoLett,WSe2
Jones,WSe2Wang,MoSe2Singh,Liu,Zhang2014,Zhu2015SciRep,Yang,ShangBiexiton,ZhangMS2,WS2Plechinger,Plechinger2016,bZu,Singh2016,substrate,CourtadeSemina,Volmer2017,Liu2019,Borghardt2020,Klein2022,Semina2022,Thomas2021ACSOmega}
observed and reported the signature of a trion in TMDCs formed from a hole
in the first valence band and electrons originating predominantly from
downwards curved conduction bands of the $K$ and/or $K^{\prime }$ valleys.
The reported trion binding energies depend on dielectric screening
(substrate versus hBN encapsulation), carrier density, and temperature.

We summarize experimentally measured trion binding energies in TMDCs, as extracted from different experimental methods of measurement (photoluminescence,
reflectance contrast, and gate-dependent optical spectroscopy).
Table~\ref{tab:trion_binding} summarizes representative experimental trion
binding energies reported for monolayer semiconducting group-VI TMDCs. Nevertheless, experimental values for monolayer TMDCs consistently fall in the
range $E_b \sim 20$--$40$~meV, as shown in Fig. \ref{TMDCtrion}, reflecting enhanced Coulomb interactions in
2D systems.


\begin{table}[h!]
\centering
\caption{Experimentally measured trion binding energies $E_b$ in monolayer TMDCs.
Binding energies are typically extracted from the energy separation between the
neutral exciton ($X^0$) and charged trion ($X^\pm$) peaks in optical spectra.
}
\label{tab:trion_binding}
\begin{tabular}{lccc}
\hline\hline
Material & Trion type & $E_b$ (meV) & Experimental reference \\ \hline

MoS$_2$
& $X^-$
& 18--20
& Mak \emph{et al.}~\cite{Mak2013NatMat} \\

MoS$_2$
& $X^-$
& $\sim$40
& Lin \emph{et al.}~\cite{Lin2014NanoLett} \\

MoSe$_2$
& $X^-$
& $\sim$30
& Ross \emph{et al.}~\cite{Ross2013NatComm} \\

MoSe$_2$
& $X^+$
& $\sim$30
& Ross \emph{et al.}~\cite{Ross2013NatComm}\footnotemark[1] \\

WS$_2$
& $X^-$
& 20--40
& Zhu \emph{et al.}~\cite{Zhu2015SciRep} \\

WS$_2$
& $X^-$
& 23--26
& Jadczak \emph{et al.}~\cite{Jadczak2017NanoLett} \\

WSe$_2$
& $X^-$
& $\sim$30
& Huang \emph{et al.}~\cite{Huang2016SciRep} \\

WSe$_2$
& $X^+$
& 25--40
& Chen \emph{et al.}~\cite{Chen2023PRB} \\

WSe$_2$
& $X^-$
& 25--40
& Chen \emph{et al.}~\cite{Chen2023PRB} \\

MoTe$_2$
& $X^+$
& $\sim$24
& Yang \emph{et al.}~\cite{Yang2015ACSNano} \\

MoTe$_2$
& $X^-$
& $\sim$27
& Yang \emph{et al.}~\cite{Yang2015ACSNano} \\

MoTe$_2$
& $X^+$
& 14--23
& Biswas \emph{et al.}~\cite{Biswas2023arXiv}\footnotemark[1] \\

MoTe$_2$
& $X^-$
& 14--23
& Biswas \emph{et al.}~\cite{Biswas2023arXiv}\footnotemark[1] \\
\hline\hline

\end{tabular}

\vspace{1mm}
\footnotetext[1]{In these experiments, $X^+$ and $X^-$
trions are not spectrally resolved with sufficient accuracy. 
}
\end{table}

\subsection{Trions in anisotropic monolayers}

Trions in phosphorene (monolayer of black phosphorus) have been investigated both experimentally and theoretically, revealing a rich and anisotropic few-body physics distinct from that of isotropic two-dimensional semiconductors.
Experimental studies have reported clear signatures of negatively charged trions through photoluminescence spectroscopy in monolayer and few-layer phosphorene, demonstrating optical tunability via photoexcitation and gating \cite{Yang2015LSA}. Subsequent experiments revealed extraordinarily large trion binding energies in few-layer phosphorene, attributed to its quasi-one-dimensional electronic structure and reduced dielectric screening \cite{Xu2016ACSNano,Yang2017CPB}. More recently, electrically controlled excitonic complexes, including trions, were observed in high-quality bilayer phosphorene devices, confirming the stability of charged excitons in gated and encapsulated samples \cite{Yoon2021PRB}.

Due to reduced dimensionality, weak dielectric screening, and strong in-plane anisotropy, excitons and trions in monolayer phosphorene exhibit pronounced quasi-one-dimensional (quasi-1D) behavior. As a result, their binding energies are substantially larger than those observed in quasi-two-dimensional quantum wells and in more isotropic 2D materials such as TMDCs. An exceptionally large trion binding energy of approximately $100~\mathrm{meV}$ was first reported in monolayer phosphorene~\cite{Yang2017CPB}, which is about two to five times larger than typical values in tungsten- or molybdenum-based TMDC monolayers. Even higher trion binding energies have been observed in few-layer phosphorene: experiments on three-layer phosphorene deposited on a SiO$_2$/Si substrate revealed ultrahigh binding energies of up to $162~\mathrm{meV}$, attributed to the formation of quasi-1D trions in the intrinsically anisotropic 2D phosphorene lattice~\cite{Xu2016ACSNano}.

Recent experimental studies have substantially refined the understanding of trion binding energies in phosphorene, revealing both large absolute values and a pronounced sensitivity to dimensionality and environment. Early photoluminescence measurements reported trion binding energies of approximately $100~\mathrm{meV}$ \cite{Yang2017CPB} in monolayer phosphorene on SiO$_2$/Si substrates~\cite{Yang2015LSA}, while few-layer systems, such as three-layer phosphorene, exhibit even larger quasi-1D  trion binding energies in the range of $160$--$180~\mathrm{meV}$~\cite{Xu2016ACSNano}. More recent experiments have provided further insight: Molas \emph{et al.}~\cite{Molas2021NatCommBP} observed trion binding energies near $120~\mathrm{meV}$ in monolayer phosphorene (the extracted binding energy of the trion in triple layer black phosphorus is
on the order of 160 meV, which is a larger value compared to the
case of the monolayer (100–120 meV) reported in Refs. \cite{Yang2015LSA,Surrente2016}) using photoluminescence spectroscopy, whereas Yoon \emph{et al.}~\cite{Yoon2021PRB} identified an additional trion resonance approximately $30~\mathrm{meV}$ below the neutral exciton under electrostatic gating. Complementary theoretical analyses and refined experimental interpretations suggest that, when referenced appropriately to specific excitonic states, the effective trion binding energy under certain conditions may be closer to $70~\mathrm{meV}$, underscoring the strong dependence of trion binding energies on dielectric screening, layer thickness, and charge density~\cite{Liang2024PCCP}. Calculations \cite{Sheng2024APhA_MB} indicate that, when substrate-induced screening is fully accounted for, the effective trion binding energy in certain few-layer configurations is reduced to approximately 70 meV, offering a more physically consistent description of the observed variability.


Related to the experimental search for trions in other anisotropic monolayers, transition-metal trichalcogenides (TMTCs, MX$_3$) have emerged over the past decade as an important class of low-dimensional materials. Among recently developed two-dimensional systems, only a limited number exhibit strong in-plane structural anisotropy, notably TMTCs such as TiS$_3$, TiSe$_3$, ZrS$_3$, and ZrSe$_3$. While TiS$_3$ is a direct-gap semiconductor in the monolayer limit, TiSe$_3$, ZrS$_3$, and ZrSe$_3$ are predicted to remain indirect-gap materials~\cite{Jin2015PCCP,Li2015Nanoscale}. The crystal structures and characteristic mechanical, electronic, optical, magnetic, and charge-density-wave transport properties of TMTCs have been comprehensively reviewed elsewhere~\cite{Dai2016WIREs,Patra2020RSC}. Experimentally, optical studies of TMTCs have revealed pronounced excitonic effects and strong photoluminescence anisotropy, including polarization-resolved photoluminescence in ZrS$_3$ and visible emission from HfS$_3$ nanoribbons, while excitonic contributions have been invoked to explain large birefringence and linear dichroism in TiS$_3$ and related compounds~\cite{Chen2023ResearchTMTC}. However, in contrast to the well-established gate-tunable trion spectroscopy in transition-metal dichalcogenide monolayers, unambiguous experimental identification of trions in monolayer  TMTCs—via controlled charge doping and clearly resolved exciton–trion splitting—remains scarce.

At present, evidence for trion states in TMTCs is primarily supported by theoretical studies, including explicit calculations of excitons and trions in monolayer TiS$_3$ \cite{Torun2018PRB_TiS3Trions}, which suggest that these complexes should be stable. Consequently, targeted gate-dependent optical experiments in high-quality, encapsulated TMTC devices represent a critical open direction for the field and would provide the necessary evidence to confirm the predicted few-body physics in these anisotropic systems.

\subsection{Trions in Xenes}
Trions are well established experimentally in direct-gap 2D semiconductors such as TMDC monolayers, where they are typically identified through gate-controlled carrier doping and a clearly resolved exciton--trion splitting in photoluminescence and/or reflectance spectra. In contrast, for Xenes---buckled group-IV monolayers such as silicene, germanene, and stanene---unambiguous experimental observation of trions remains limited. A key practical limitation is that these materials are commonly realized epitaxially on metallic templates, where strong hybridization and screening can substantially modify (or suppress) intrinsic optical responses and hinder clean charge-tunable exciton spectroscopy~\cite{Zhao2016PR,Scalise2018NanoRes,Ochapski2022OpenPhys}. As a result, trions in Xenes have been investigated predominantly at the theoretical level, including predictions of electrically tunable trion binding energies and related many-body effects in silicene, germanene, and stanene under perpendicular electric fields~\cite{Kezerashvili2024PRB,Kezerashvili2024arXivXeneTrions}. While functionalized Xenes (e.g., germanane and related derivatives) exhibit measurable photoluminescence and excitonic photophysics, systematic gate-dependent studies establishing charged-exciton resonances in the same definitive manner as TMDCs are still largely missing~\cite{Cinquanta2022Germanane,Kupchak2024GermaneneFunc}. These considerations indicate that high-quality, encapsulated, charge-tunable Xene devices on insulating substrates would be essential for a definitive experimental trion spectroscopy in this material family.

Theoretical studies of excitons and magnetoexcitons have demonstrated that the binding energies and optical properties of both direct and indirect excitonic states can be strongly tuned by external electric and magnetic fields~\cite{Kezerashvili2018PRBRoma,Kezerashvili2021PRBRoma}. These works showed that the interplay between field-induced band-structure modifications and the RK interaction leads to significant control of excitonic spectra, including Rydberg states and diamagnetic coefficients, suggesting potential applications in electrically and magnetically tunable optoelectronic devices. Similar approaches can be extended to the study of trions in Xenes.



\section{Theoretical approaches and formalism}
\label{Sect_6}

This section summarizes the general low-energy formalism used to describe
Mott--Wannier trions in two-dimensional materials within the framework of the non-relativistic approach. We introduce the effective
three-body Hamiltonian, employ mass-scaled Jacobi coordinates to separate the
center-of-mass ($c.m.$) motion, and derive the Schr\"odinger equation governing the
relative motion of the three charged particles. Throughout this work,
Mott--Wannier trions in a two-dimensional semiconductor are modeled within the
isotropic effective-mass approximation. The electrostatic interactions between
electrons and holes are assumed to be isotropically screened by the static
dielectric permittivity of the crystal. A trion is treated as a
three-particle bound state: a negatively charged trion, $X^{-}$, consists of
two electrons and one hole, whereas a positively charged trion, $X^{+}$,
consists of two holes and one electron. Within this framework, the non-relativistic Mott--Wannier trion Hamiltonian in a
2D plane is given by
\begin{equation}
H = -
\sum_{i=1}^{3}\frac{\hbar^{2}}{2m_{i}}\nabla_{i}^{2}
+ \sum_{i<j}^{3}
V_{ij}\!\left(|\mathbf{r}_{i}-\mathbf{r}_{j}|\right),
\label{eq:TrionHamiltonian}
\end{equation}
where $m_i$ and $\mathbf{r}_i$ denote the effective mass and the in-plane
position vector of the $i$th particle, respectively. Electrons and holes are
treated within the band effective-mass approximation. The interaction
potential $V_{ij}$ describes the screened Coulomb interaction between charged
carriers in a 2D material. Depending on the level of approximation, it can be modeled by
the Rytova--Keldysh (\ref{RKeldysh}), modified
Kratzer-type ~(\ref{Kr25}) potentials, or other effective screened interaction
forms such as those introduced in Ref.~\cite{Tuan2018PRB}.
Since the Hamiltonian (\ref{eq:TrionHamiltonian}) is invariant under two-dimensional rotations,
the total orbital angular momentum $J$ is conserved and therefore a good quantum number.

The corresponding Schr\"odinger equation for the three-particle system reads
\begin{equation}
\left[
-\sum_{i=1}^{3}\frac{\hbar^{2}}{2m_{i}}\nabla_{i}^{2}
+ \sum_{i<j}^{3}
V_{ij}\!\left(|\mathbf{r}_{i}-\mathbf{r}_{j}|\right)
\right]
\Psi(\mathbf{r}_{1},\mathbf{r}_{2},\mathbf{r}_{3})
=
\mathcal{E}\,\Psi(\mathbf{r}_{1},\mathbf{r}_{2},\mathbf{r}_{3}).
\label{eq:SchrodingerFull}
\end{equation}

In crystalline solids, the effective mass of charge carriers is determined by
the local curvature of the electronic band dispersion evaluated at a band
extremum (valley),
$\left(m^{*}_{\alpha}\right)^{-1}
=\frac{1}{\hbar^{2}}
\left.
\frac{\partial^{2}E(\mathbf{k})}{\partial k_{\alpha}^{2}}
\right|_{\mathbf{k}=\mathbf{k}_{0}}$  ~\cite{AshcroftMermin,YuCardona},
where $\mathbf{k}_{0}$ denotes the position of the valley in the Brillouin
zone and $\alpha$ labels the Cartesian components. Consequently, when
different valleys correspond to inequivalent points in the Brillouin zone, their band curvatures—and therefore the corresponding effective masses—can differ. In two-dimensional materials, this leads to a valley dependence of the electron
and hole effective masses that are both material- and band-specific.

In monolayer TMDCs, the K and K$^{\prime}$
valleys are related by time-reversal symmetry and thus possess identical
effective masses for electrons and holes~\cite{Xiao2012PRL,Kormanyos,
Manzeli2017}. By contrast, other valleys, such as the Q valleys in the
conduction band or the $\Gamma$ point in the valence band, are inequivalent
and often characterized by distinct and anisotropic effective masses, which
can become relevant at elevated carrier densities, under applied strain, or
in multilayer structures~\cite{Kormanyos}.

The valley dependence of effective masses has important consequences for the fine structure of trions in two-dimensional materials. For intravalley trions formed from carriers near the K and K$^{\prime}$ valleys, the equality of effective masses ensures valley-degenerate binding energies in the absence of external fields. However, for intervalley trions or trions involving carriers from inequivalent valleys (e.g., K–Q or $\Gamma$–K configurations), differences in effective masses modify the reduced masses entering the three-body Hamiltonian (\ref{eq:TrionHamiltonian}). This leads to distinct trion binding energies, altered level ordering, and valley-dependent exchange splittings~\cite{model,Mostaani2017,Tempelaar2019}. Such effects contribute to the experimentally observed richness of trion spectra and their sensitivity to doping, strain, and dielectric environment.

\begin{figure}[H]
\centering
\includegraphics[width=12cm]{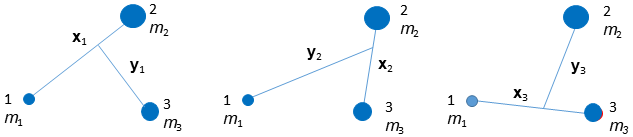}
\caption{Schematic illustration of the mass-scaled Jacobi coordinates set for three nonidentical particles forming a trion.}
\label{fig:Jacobi}
\end{figure}

\noindent
Therefore, in general, the particle masses $m_i$ appearing in Eqs.~(\ref{eq:TrionHamiltonian}) and (\ref{eq:SchrodingerFull}) are not identical, reflecting the possible
differences between electron and hole effective masses and, in some cases,
their valley or band dependence. As a consequence, an explicit separation of
the $c.m.$ motion from the internal dynamics of the three-body system
is required. To this end, we introduce mass-scaled Jacobi coordinates shown in Fig. \ref{fig:Jacobi}. For a
given partition $i$ ($i \neq j \neq k = 1,2,3$), these coordinates are defined as
\begin{eqnarray}
\mathbf{x}_{i} &=&
\sqrt{\frac{m_{j}m_{k}}{(m_{j}+m_{k})\mu}}
(\mathbf{r}_{j}-\mathbf{r}_{k}), \notag \\
\mathbf{y}_{i} &=&
\sqrt{\frac{m_{i}(m_{j}+m_{k})}{(m_{1}+m_{2}+m_{3})\mu}}
\left(
\mathbf{r}_{i}-\frac{m_{j}\mathbf{r}_{j}+m_{k}\mathbf{r}_{k}}{m_{j}+m_{k}}
\right),
\label{eq:Jacobi} \\
\mathbf{R} &=&
\frac{m_{1}\mathbf{r}_{1}+m_{2}\mathbf{r}_{2}+m_{3}\mathbf{r}_{3}}{M},
\qquad
M = m_{1}+m_{2}+m_{3}, \notag
\end{eqnarray}
where $\mathbf{R}$ is the $c.m.$ coordinate  and
\begin{equation}
\mu = \sqrt{\frac{m_{1}m_{2}m_{3}}{m_{1}+m_{2}+m_{3}}}
\end{equation}
is the three-particle effective mass.

After the transformation to Jacobi coordinates, the $c.m.$ motion
separates, and the Schr\"odinger equation for the relative motion of the
three-body system becomes
\begin{equation}
\left[
-\frac{\hbar^{2}}{2\mu}
\left(
\nabla_{\mathbf{x}_{i}}^{2} + \nabla_{\mathbf{y}_{i}}^{2}
\right)
+ \sum_{i<j}^{3} V_{ij}
- \mathcal{E}
\right]
\Psi(\mathbf{x}_{i},\mathbf{y}_{i}) = 0.
\label{eq:RelativeSchrodinger}
\end{equation}

\noindent
Equation~(\ref{eq:RelativeSchrodinger}) governs the internal dynamics of
two-dimensional trions and forms the starting point for further analysis,
including solutions based on a variety of few-body
approaches discussed in the following subsections.



\subsection{Deterministic Few-Body Methods}
Below, we consider the variational method, stochastic variational method, exact (direct) diagonalization, and Gaussian expansion method. The common foundation of these approaches is the variational principle, which guarantees that any properly normalized trial wave function yields an upper bound to the exact ground-state energy. By refining the functional form of the wave function or enlarging the basis set, these deterministic approaches systematically converge toward the exact solution within the effective-mass approximation. In two-dimensional materials, where the Rytova--Keldysh potential introduces nontrivial screening effects and strong carrier correlations, deterministic frameworks are essential for achieving high numerical precision. In these methods, the Schr\"odinger equation is solved by expanding the wave function in a suitable basis and determining the eigenvalues through variational optimization or matrix diagonalization. In the following subsections, we review four deterministic approaches applied to trions in 2D materials: the variational method, the stochastic variational method with explicitly correlated Gaussians, exact diagonalization, and the Gaussian expansion method.


\subsubsection{Variational method}

The variational method (VM) is a powerful and systematically improvable approach for obtaining accurate approximations to bound-state solutions of few-body quantum systems. Its foundation lies in the Rayleigh--Ritz principle, according to which the expectation value of the Hamiltonian operator,
\begin{equation}
\mathcal{E}[\Psi] = \frac{\langle \Psi | \hat{H} | \Psi \rangle}{\langle \Psi | \Psi \rangle},
\end{equation}
provides an upper bound to the exact ground-state energy for any normalized trial wave function $\Psi$ that satisfies the required symmetry and boundary conditions.

In the framework of VM the trial wave function is constructed as a linear superposition of basis states,
\begin{equation}
\Psi = \sum_{i=1}^{N} c_i \, \Phi_i,
\end{equation}
where $\{\Phi_i\}$ denotes a chosen set of basis functions and $\{c_i\}$ are variational coefficients to be determined. Substituting this expansion into the functional $E[\Psi]$ and minimizing with respect to the coefficients $c_i$ leads to a generalized eigenvalue problem,
\begin{equation}
\sum_{j=1}^{N} \left( H_{ij} - \mathcal{E} \, S_{ij} \right) c_j = 0,
\end{equation}
where
\begin{equation}
H_{ij} = \langle \Phi_i | \hat{H} | \Phi_j \rangle,
\qquad
S_{ij} = \langle \Phi_i | \Phi_j \rangle
\end{equation}
are the Hamiltonian and overlap matrices, respectively.

The accuracy of the variational solution depends on the completeness and flexibility of the chosen basis. Increasing the basis size $N$ or introducing additional nonlinear variational parameters systematically improves the approximation and ensures convergence of the energy from above toward the exact value. In practice, convergence is monitored by examining the stability of the eigenvalues with respect to variations of the basis parameters and by verifying the saturation of relevant observables.

In addition to providing reliable ground-state energies, the variational framework allows for the computation of structural properties such as root-mean-square radii, correlation functions, and probability densities, which can be evaluated directly using the optimized wave function. The method is particularly well suited for few-body systems, where high numerical precision can be achieved with a relatively moderate basis size or trial wave functions.

The seminal 2013 paper by Berkelbach, Hybertsen, and Reichman \cite{Berkelbach2013} provided the first comprehensive variational calculations for trions in TMDC monolayers using the Rytova–Keldysh potential. In Ref.~\cite{Chang2021Trion}, high-precision variational calculations of trions in TMDCs were performed using trial wave functions constructed from two-dimensional Slater-type orbitals in the form introduced in Ref.~\cite{Wu2019or}. The variational trion wave function was expressed as a linear combination of products of Slater-type orbitals. The authors in Ref.~\cite{Semina2022Suris} also provided a comprehensive
review of theoretical studies on the structure and binding energies of
localized excitons and trions in nanosystems within the variational
framework (see also references therein).



\subsubsection{Stochastic Variational Method}
The stochastic variational method (SVM) provides high-precision solutions of few-body quantum systems by expanding
the many-body wave function in a flexible basis of explicitly
correlated Gaussians (ECGs) \cite{SuzukiVarga1998}. The method has been extensively
validated for atomic, molecular, nuclear, and excitonic few-particle systems and is particularly well suited for treating both ground and excited states of excitons, trions, and biexcitons in reduced dimensions.

Within the effective-mass approximation, for trions consisting of three particles confined to a 2D plane described by the nonrelativistic
Hamiltonian (\ref{eq:TrionHamiltonian}), the variational wave function is expanded in a basis of
ECGs~\cite{Mitroy2013},
\begin{equation}
\Psi
=
\sum_{k=1}^{K} c_k \,
\mathcal{A}
\left\{
\chi^{S}_{\mathrm{spin}}
\left[
\prod_{i=1}^{3} \xi_{m_i}(\mathbf{r}_i)\,
\exp\!\left(
-\frac{1}{2}
\sum_{i,j=1}^{3}
A_{ij}\,\mathbf{r}_i \cdot \mathbf{r}_j
\right)
\right]
\right\},
\label{eq:ECG}
\end{equation}
with the angular prefactor
\begin{equation}
\xi_{m}(\mathbf{r}) = (x + i y)^{m},
\label{eq:xi_m}
\end{equation}
where 
$\mathcal{A}$ is the antisymmetrization operator acting on
identical fermions, $\chi^{S}_{\mathrm{spin}}$ is 
spin function coupling the electrons’ and holes’
spins to total spin $S$, and the set of nonlinear parameters
$\{A_{ij}\}$ defines the Gaussian widths and correlations. The nonlinear variational parameters are optimized using the stochastic
variational method~\cite{SuzukiVarga1998,Mitroy2013}.
The total angular momentum in two dimensions is
$\mathfrak{L} = \sum_{i=1}^{3} \mathfrak{l_i}$
 with projection $\mathfrak{M} = \mathfrak{m}_1 + \mathfrak{m}_2 + \mathfrak{m}_3$, the integers $\mathfrak{m}_i$ determine the angular
momentum projection.
The spatial wave function in Eq.~(\ref{eq:ECG}) is coupled with the
spin function $\chi_{S M}^S$ to form the complete trial wave function.

The ECG basis efficiently captures both strong short-range correlations and extended spatial structures, making it well suited for weakly bound and excited excitonic complexes. The nonlinear parameters are optimized stochastically by retaining only basis functions that lower the total energy, allowing the basis to adapt to the dominant correlations while maintaining numerical stability. ECGs are widely used in nuclear and atomic physics, and quantum chemistry~\cite{SuzukiVarga1998} due to: i. the analytical availability of matrix elements; ii. high variational flexibility; iii. straightforward treatment of permutation symmetry; iv. simple coordinate transformations.

Material-dependent parameters, such as effective masses and
screening lengths, are taken from \emph{ab initio}-based
parameterizations known to reproduce exciton binding energies in
good agreement with Bethe--Salpeter calculations. For simplicity,
and in accordance with experimental observations, calculations
are often performed assuming equal electron masses in $X^-$ and hole masses in $X^+$,
which ensures charge-conjugation symmetry between positive and
negative trions without affecting the qualitative behavior of the
binding energies.
The calculated exciton binding energies are found to be in good
agreement with experimental values, typically on the order of
several hundred millielectronvolts in monolayer transition metal
dichalcogenides. Trions \cite{Kidd2016,Varga2020} and biexciton \cite{VargaNano2015} binding energies are also well reproduced
within this framework, demonstrating the reliability of the
effective-mass model combined with the SVM for three-body
excitonic complexes.



\subsubsection{Exact Diagonalization Method}

The exact diagonalization (ED) method, also referred to as direct diagonalization, provides a nonperturbative approach to few-body excitonic complexes by representing the Hamiltonian in a finite basis and solving the resulting matrix eigenvalue problem numerically. In the context of trions in 2D semiconductors, the method is typically implemented within the effective-mass approximation, employing screened Coulomb interactions modeled via the Rytova–Keldysh potential to account for non-local screening effects.

To make the problem tractable, after separating the $c.m.$ motion, the relative
three-body wave function $\Psi(\mathbf{x},\mathbf{y})$, in Jacobi coordinates, is expanded in a finite basis set of dimension $N_b$:
\begin{equation}
\Psi(\mathbf{x},\mathbf{y}) = \sum_{n=1}^{N_b} c_n \Phi_n(\mathbf{x},\mathbf{y}).
\end{equation}
Typical choices for the basis functions $\{\Phi_n\}$ include 2D harmonic-oscillator states, plane waves, or specialized Gaussian-type orbitals. The Hamiltonian matrix elements, $H_{mn} = \langle \Phi_m | \hat H | \Phi_n \rangle$, are computed explicitly, leading to a finite-dimensional eigenvalue problem:
\begin{equation}
\mathbf{H} \mathbf{c} = \mathcal{E} \mathbf{c}.
\label{EVP}
\end{equation}
Diagonalization of the matrix eigenvalue problem (\ref{EVP}) yields the discrete energy spectrum, providing direct access to both ground and excited states, as well as the corresponding wave functions. The method is numerically exact within the chosen finite basis. However, exact diagonalization (ED) suffers from the ``curse of dimensionality,'' since the basis size required for convergence grows rapidly—typically exponentially—with the number of particles $N$, which typically limiting the method to $N \le 4$.

If the system possesses rotational symmetry, as is commonly the case for the ground state of a trion, the total angular momentum can be separated. In this situation, the wave function may be decomposed into angular and radial parts, and the basis expansion is performed over the radial component only. The resulting basis functions then depend solely on the magnitudes of the Jacobi coordinates and the relative angle between them.

The ED approach has been applied to exciton--electron systems in configuration space~\cite{Fey2020}
and to the study of trions in TMDC using a momentum-space formulation~\cite{Kumar2025}. Notably, ED provides a vital deterministic benchmark for Monte Carlo approaches such as diffusion Monte Carlo 
and path-integral Monte Carlo.

\subsubsection{Gaussian Expansion Method}

Over the past two decades, the Gaussian expansion method (GEM),
which is based on a systematic expansion of the wave function
in Gaussian basis functions, has been widely and successfully
applied to few-body systems in nuclear physics for the precise
description of three- and four-particle systems; see, for example,
Refs.~\cite{Hiyama2000,Kamada2001,Hiyama2003,Hiyama2004,Hiyama2010,Hiyama2012}.
The GEM is a deterministic
variational technique for solving few-body Schr\"odinger equations
in configuration space. Most reccently CEM was applied to investigate the properties of trions in TMDCs monolayers \cite{Tenorio2026GEM2D}. After separating the $c.m.$ motion, the trion wave function is written
in Jacobi coordinates and expanded over a finite set of Gaussian basis states,
often organized into \emph{rearrangement channels} (different Jacobi partitions)
to efficiently capture alternative correlation topologies (e.g., an exciton plus
a spectator carrier). In a compact form, the variational ansatz for a state with total orbital angular momentum $J$ can be written as \cite{Tenorio2026GEM2D}
\begin{equation}
\Psi^{J}(\mathbf{x},\mathbf{y})
=
\sum_{i}\sum_{\alpha_i}
A^{i}_{\alpha_i}\,
\Phi^{i}_{\alpha_i}(\mathbf{x}_i,\mathbf{y}_i),
\label{eq:GEM_expansion}
\end{equation}
where $(\mathbf{x}_i,\mathbf{y}_i)$ are the Jacobi coordinates in partition $i$,
$\alpha_c$ labels the quantum numbers and nonlinear width parameters of the
Gaussian basis, and $A^{i}_{\alpha_i}$ are linear expansion coefficients \cite{Tenorio2026GEM2D}.

In each channel, the basis functions are constructed from Gaussian radial factors
(with widths chosen in a geometric progression to cover both short- and long-range
length scales) multiplied by angular functions that carry the total angular
momentum $J$. Substituting Eq.~\eqref{eq:GEM_expansion} into the Schr\"odinger
equation and applying the Rayleigh--Ritz variational principle leads to a
generalized secular equation for eigenvalue problem with Hamiltonian in Jacobi coordinates from Eq. (\ref{eq:RelativeSchrodinger}) and corresponding overlap matrices.

Diagonalization of secular equation 
yields the discrete spectrum and
wave functions. Because GEM is variational, the computed energies are upper
bounds to the exact eigenvalues within the assumed Hamiltonian, and convergence
is systematically improved by enlarging the basis and/or the set of channels.

Although both GEM and the SVM employ Gaussian basis functions, their philosophies differ fundamentally. In GEM, the nonlinear Gaussian width parameters are fixed in advance using a deterministic geometric progression, and convergence is achieved by increasing the basis size across multiple rearrangement channels. In contrast, SVM uses explicitly correlated Gaussians whose nonlinear parameters are optimized stochastically through iterative energy minimization. This adaptive optimization allows SVM to capture interparticle correlations more efficiently with fewer basis functions, often achieving benchmark-level precision. GEM, however, is computationally simpler and fully deterministic, whereas SVM requires nonlinear optimization but provides greater flexibility in describing correlated few-body dynamics.

In the framework of  deterministic approaches, the trion binding energy is extracted by comparing the trion ground-state energy to that of the isolated exciton:
\begin{equation}
\mathcal{E}_b(X^\pm) = \mathcal{E}(X) - \mathcal{E}(X^\pm),
\label{eq:Eb_det}
\end{equation}
where $\mathcal{E}(X)$ and $\mathcal{E}(X^{\pm})$ are obtained independently using the same basis framework with the exciton and trion Hamiltonians, respectively, to ensure consistent error cancellation. These methods serve as the primary validation for diffusion Monte Carlo and path-integral Monte Carlo results, particularly in the few-body limit where basis convergence is attainable.

To conclude, deterministic approaches—most notably the SVM, GEM, and ED, provide high-precision benchmarks for few-body excitonic complexes. By representing the Hamiltonian in a basis set and solving the resulting equations, these methods bypass the statistical noise inherent in Monte Carlo techniques.



\subsection{Quantum Monte Carlo methods}

Quantum Monte Carlo (QMC) \cite{Hammond1994,Foulkes2001} is an umbrella term for stochastic numerical methods used to solve quantum many-body problems by Monte Carlo sampling. It includes several approaches that differ in formulation and target observables. Diffusion Quantum Monte Carlo (DMC) \cite{CeperleyAlder1980, Foulkes2001} is a zero-temperature QMC method that projects out the exact ground state by evolving a trial wave function in imaginary time. For fermionic systems, it relies on the fixed-node approximation to mitigate the fermion sign problem by imposing nodal constraints, as the problem remains fundamentally unsolved and has been shown to be computationally NP-hard \cite{Troyer2005} for general fermionic systems.

In contrast, path-integral Monte Carlo (PIMC) \cite{GlimmJaffe1987,NegeleOrland1988,Ceperley1995} is a finite-temperature QMC method based on the imaginary-time path-integral representation of the density matrix, which samples thermal ensembles without an explicit trial wave function and provides direct access to thermodynamic quantities. While DMC is most effective for zero-temperature fermionic systems, PIMC naturally includes thermal effects but becomes severely limited by the fermion sign problem at low temperatures.



\subsubsection{Diffusion Monte Carlo for 2D trions 
}

Diffusion Quantum Monte Carlo is a stochastic projector method designed to extract the ground-state properties of many-body systems. The method is based on the observation that the Schrödinger equation in imaginary time, $\tau=it$, takes the form of a generalized diffusion equation. For a system described by the Hamiltonian (\ref{eq:TrionHamiltonian}), the evolution of a trial wave function $\Psi_T$ is given by:

\begin{equation}
-\frac{\partial \Psi(\mathbf{R},\tau)}{\partial \tau}
= \left(\hat H - \mathcal{E}_T\right)\Psi(\mathbf{R},\tau),
\label{eq:imagtime}
\end{equation}
where $\mathbf{R}=(\mathbf{r}_1,\mathbf{r}_2,\mathbf{r}_3)$ represents the $9$-dimensional configuration space of the particles and $E_T$ is an energy offset used to control the walker population.
As $\tau\to\infty$, the components of the trial wave function corresponding to excited states decay exponentially faster than the ground state, $\Psi(\mathbf{R},\tau)$ approaches the lowest-energy eigenstate consistent with the imposed symmetry.

To reduce variance, one uses importance sampling with a (real, positive) trial wave function $\Psi_T(\mathbf{R})$
and propagates the mixed distribution $f(\mathbf{R},\tau)=\Psi_T(\mathbf{R})\Psi(\mathbf{R},\tau)$:
\begin{equation}
-\frac{\partial f}{\partial \tau}
= -\frac{1}{2}\nabla_{\mathbf{R}}^2 f
  + \nabla_{\mathbf{R}}\!\cdot\!\left[\mathbf{v}_D(\mathbf{R}) f\right]
  + \left(\mathcal{E}_L(\mathbf{R})-\mathcal{E}_T\right) f,
\label{eq:fp}
\end{equation}
where $\nabla_{\mathbf{R}}$ is the gradient in the full configuration space,
\begin{equation}
\mathbf{v}_D(\mathbf{R})=\nabla_{\mathbf{R}} \ln \Psi_T(\mathbf{R})
\label{eq:drift}
\end{equation}
is the drift velocity, and
\begin{equation}
\mathcal{E}_L(\mathbf{R})=\frac{\hat H \Psi_T(\mathbf{R})}{\Psi_T(\mathbf{R})}
\label{eq:localE}
\end{equation}
is the local energy. Equation~\eqref{eq:fp} has the form of a drift--diffusion process with a branching term
proportional to $E_L-E_T$ \cite{Mayers2015,Szyniszewski2017,Mostaani2017}.

In the DMC calculations, usually
a Slater--Jastrow trial wave function was employed,
\begin{equation}
\Psi_T(\mathbf{R})
=
\mathcal{A}\left\{
\chi^{S}_{\mathrm{spin}}
\left[
\prod_i \phi_i(\mathbf{r}_i)
\right]
\exp\!\left(
\sum_{i<j} u(r_{ij})
\right)\right\},
\end{equation}
where $\mathcal{A}$ enforces antisymmetry for identical fermions,
$\phi_i$ are single-particle orbitals,
and $u(r_{ij})$ is a two-body Jastrow correlation factor
that satisfies the appropriate Coulomb cusp conditions.

In the context of TMDC, DMC has proven instrumental for determining the stability of complex many-body states. Studies ~\cite{Mayers2015,Szyniszewski2017,Mostaani2017,Mostaani2017_DMC,Mostaani2023} utilize DMC to calculate the binding energies of excitonic complexes. These calculations typically involve a model Hamiltonian that accounts for the screened RK potential (\ref{RKeldysh}).

Trions binding energy are computed relative to dissociation into an exciton plus a free carrier by analogously as given in Eq. (\ref{eq:Eb_det}), where $\mathcal{E}(X)$ and $\mathcal{E}(X^\pm)$ are the total ground-state energies obtained from DMC for the exciton and trion Hamiltonians, respectively \cite{Mayers2015,Szyniszewski2017,Mostaani2017,Mostaani2017_DMC,Mostaani2023}.
In addition to binding energies, DMC yields statistically exact pair-correlation functions and contact densities,
which can be used to parameterize short-range exchange or contact interactions perturbatively
\cite{Szyniszewski2017,Mostaani2017,Mostaani2017_DMC,Mostaani2023}.

 By employing DMC, one can bypass the limitations of variational methods and obtain high-accuracy results that account for the strong correlation between carriers in monolayer TMDCs \cite{Mayers2015}. Furthermore, this approach allows for the investigation of how the binding energy scales with the effective mass ratio and the dielectric environment \cite{ Mostaani2017,Mostaani2017_DMC,Mostaani2023}, providing a benchmark for experimental photoluminescence data.


\subsubsection*{ Path-integral Monte Carlo for 2D trions}

The path-integral Monte Carlo method provides a nonperturbative framework for treating strongly correlated few-body quantum systems as thermal systems at a finite temperature.
The approach is based on the real-time path-integral formulation of the density matrix
introduced by Feynman~\cite{Feynman1948,FeynmanHibbs1965}.
Then the central idea is that the few-body system, which has no thermal properties,
is translated {\it formally} into a grand-canonical ensemble at a non-zero temperature $T$
by an analytic continuation of the real time $t$ to the imaginary time $-it$, also known as
Wick rotation.  The advantage of this translation is that the thermal system can be treated
with standard techniques of Monte-Carlo integration.
In contrast to diffusion DMC, which projects the ground-state wave function $\Psi(\mathbf{R})$ in imaginary time, PIMC is formulated directly in terms of the finite-temperature density matrix and does not rely on a trial ground-state wave function.

A central quantity in PIMC is the canonical partition function
\begin{equation}
Z = \mathrm{Tr}\, e^{-\beta \hat{H}},
\end{equation}
where $\hat{H}$ is the trion Hamiltonian and
$\beta = 1/(k_B T)$, with $k_B$ denoting the Boltzmann constant
and $T$ the temperature. In coordinate representation, the trace is expressed as
\begin{equation}
Z = \int d\mathbf{R}\; \langle \mathbf{R} | e^{-\beta \hat{H}} | \mathbf{R} \rangle ,
\end{equation}
where $\mathbf{R} = (\mathbf{r}_1,\mathbf{r}_2,\mathbf{r}_3)$ denotes the set of particle coordinates.

To clarify the origin of the imaginary-time formulation used in PIMC,
one may start from the real-time evolution operator describing the
dynamics of the system over the time interval $[0,t]$,
\begin{equation}
U(t)=e^{-i\hat{H}t/\hbar}.
\end{equation}
Applying the Wick rotation $t \rightarrow -i\beta\hbar$ transforms the
real-time evolution into imaginary-time evolution, yielding the
thermal distribution operator (Boltzmann density operator) $e^{-\beta \hat{H}}$, which
appears in the partition function. To calculate $Z$, we use
the Trotter decomposition \cite{Trotter1959,Suzuki1976}, in which
the density matrix is factorized into $M$ imaginary-time slices
\begin{equation}
e^{-\beta \hat{H}} =\lim_{M\to\infty}
\left(e^{-\Delta\tau \hat{H}}\right)^M,
\qquad
\Delta\tau = \beta/M .
\end{equation}
For a finite $M$ we can write the product of operators $e^{-\Delta\tau \hat{H}}$
in coordinate representation
$\langle \mathbf{R} | e^{-\Delta\tau \hat{H}} | \mathbf{R'} \rangle$. After inserting
this matrix product into $Z$ we get the path integral
\begin{equation}
Z =\lim_{M\to\infty}\int\prod_{k=1}^{M} d\mathbf{R}_k
\exp\left[
-\sum_{k=1}^{M}
\left(\frac{m}{2\hbar^2\Delta\tau}
|\mathbf{R}_k-\mathbf{R}_{k+1}|^2
+
\Delta\tau V(\mathbf{R}_k)
\right)
\right].
\end{equation}
This mapping transforms the quantum problem into an equivalent
classical system of three interacting ring polymers, each consisting
of $M$ imaginary-time beads connected by harmonic springs arising
from the kinetic-energy contribution.

In a general treatment of three charged fermions of different masses,
the constituting particles of the trion can be treated as distinguishable, and
exchange permutations are therefore neglected. Thus, the partition
function is written in the Boltzmann form without antisymmetrization
of the density matrix. The absence of permutation operators indicates
that exchange effects between identical particles are ignored, while
the quantum nature of the particles is still fully retained through the
path-integral formulation. Corresponding expressions for fermionic
PIMC can be found in Ref.~\cite{Filinov2015PIMC}, where permutations
between identical particles are explicitly included.

PIMC has been applied extensively to excitonic complexes
in two-dimensional semiconductors, including excitons,
trions, and biexcitons, where it captures strong Coulomb
correlations and finite-temperature effects.
Representative applications to monolayer transition-metal
dichalcogenides and related systems include
Refs.~\cite{Kylanpaa2015,Velizhanin2015PIMC}. PIMC also applied to excitonic complexes in quantum wells. Filinov, Bonitz, and Lozovik \cite{Filinov2004QW} performed one of the first first-principles path-integral Monte Carlo studies of strongly correlated excitonic complexes—including excitons, trions, and biexcitons—in semiconductor quantum wells, obtaining accurate binding energies and correlation properties as functions of quantum-well width.

Thermodynamic observables, such as the internal energy and pair-correlation functions, are obtained as ensemble averages over these imaginary-time paths. PIMC has been extensively applied to excitons, trions, and biexcitons in quantum dots, see, for example application for trions in quantum dots  \cite{Filinov2003JPAQ}, where it accurately captures strong Coulomb correlations and finite-temperature effects. Later works used improved PIMC algorithms to treat larger electron clusters in quantum dots and analyze Wigner-molecule formation and correlation effects \cite{WeissBonitz2005}.

Authors \cite{Filinov_Varga2017PRB} report binding energies of trions and provide a direct, head-to-head comparison between the SVM and PIMC results. The authors demonstrated that at the zero-temperature limit, $T\to 0$, the results of the PIMC calculations converge with the SVM results. This was a critical "sanity check" for the community, proving that both stochastic wavefunction optimization within the SVM framework and stochastic sampling of the density matrix within PIMC yield the same binding energies for trions and biexcitons.
Additionally, the combined density functional theory and PIMC study by Kylänpää and Komsa \cite{Kylanpaa2015} provides further PIMC binding-energy results for exciton complexes.

Unlike variational approaches, PIMC does not require an explicit trial wave function and is therefore free of variational bias. However, for fermionic systems it suffers from the fermion sign problem, which results from the antisymmetrization of the density matrix and limits simulations to small particle numbers or moderate temperatures. Despite this limitation, PIMC provides an important benchmark for zero-temperature variational and diffusion Monte Carlo methods and plays a central role in validating theoretical descriptions of excitonic complexes in two-dimensional semiconductors.

DMC calculations of Mott–Wannier trions in 2D systems have been carried out primarily for monolayer TMDC and idealized 2D models, Refs.~\cite{Mayers2015,Szyniszewski2017,Mostaani2017,Mostaani2023}, as well as for indirect trions in coupled quantum wells modelled by two-dimensional
bilayers Ref.~\cite{Witham2018}.



\subsection{Faddeev equations approach for trions}

Because trions are intrinsically three-body systems, their theoretical description
requires a genuine three-body treatment beyond effective two-body models.
In particular, the pairwise nature of interactions and the mass asymmetry between electrons and holes prevents an exact separation into
independent subsystems. A natural and rigorous framework for addressing such
problems is provided by the Faddeev equations, which allow the systematic
decomposition of the total three-body wave function into components associated
with pairwise interactions.

Within the Faddeev formalism, the total trion wave function $\Psi$ can be expressed in terms of Faddeev components $\psi_{ij}$ as:

\begin{equation}
\Psi=\sum_{i<j}^{3}\psi_{ij}.
\end{equation}

\noindent
The component $\psi_{ij}$ corresponds to a specific interacting pair $(ij)$, with $i\neq j \neq k = 1,2,3$.  Each Faddeev component satisfies an
integral \cite{Fad,Fad1} or differential \cite{Merkuriev1976ThreeBody,FM} equation that explicitly accounts for the
interaction within the selected pair while incorporating the influence of the
third particle through coupling to the remaining components. The number of Faddeev components $\psi_{ij}$ equals the number of distinct binary interactions. These equations constitute a set of three Faddeev equations for a three-body problem, read

\begin{equation}
\psi_{ij}= (\mathcal{E}-\hat{H}_{0}) ^{-1}\hat{V}_{ij}  \Psi,
\end{equation}
where $\hat{H}_0=-\sum_{i=1}^{3}\frac{\hbar^{2}}{2m_{i}}\nabla_{i}^{2}$ is the  Hamiltonian of three non-interacting particles and $\hat{V}_{ij}$ denotes the interaction between particles $i$ and $j$.

The advantage of the Faddeev approach lies in its ability to treat all
pairwise interactions on an equal footing and to properly account for
correlation effects that are essential for binding in three-body
systems. In particular, it avoids double counting of interaction terms and
ensures the correct asymptotic behavior of the wave function. This makes the
formalism especially suitable for studying trion binding energies and
structural properties in low-dimensional systems, where correlation effects
are enhanced.

In practical applications to semiconductor trions, the Faddeev equations may
be formulated either in configuration space or in momentum space, depending
on the choice of interaction potential and numerical strategy. In bulk
materials, configuration-space differential Faddeev equations with Coulomb
interactions were first successfully employed to study trions in
Refs.~\cite{Filikhin2018,Filikhin2018Nano}.

In contrast, Ref.~\cite{Mohseni2023PRB} developed a momentum-space Faddeev
formalism to investigate trions in semiconductor layered materials, treating
three charged particles confined to two dimensions. In that work, the authors
solved the coupled Faddeev integral equations in momentum space for both a short-range separable Yamaguchi potential and the long-range Rytova--Keldysh interaction,
with applications to trions in a MoS$_2$ monolayer.

For trions, symmetry considerations play a crucial role. In the case of identical effective masses for electrons or holes, a
negative trion $X^{-}$, consisting of two identical electrons and one hole,
the total wave function must be antisymmetric under exchange of the two
electrons. This constraint reduces the number of independent Faddeev
components and leads to a coupled set of two differential equations \cite{KezJPG2016,KezPRD2020,KezJPG2024,FKVPRD2024,FRV25,RKFilikhin2025PRD} with well-defined symmetry
properties. An analogous consideration applies to positive trions $X^{+}$,
where the two holes are identical fermions.

The Faddeev framework thus provides a rigorous theoretical foundation for the
description of trions, complementing variational, hyperspherical, and quantum
Monte Carlo approaches, and offering valuable insight into the role of
three-body correlations in excitonic complexes.

\subsection{Hyperspherical harmonics formalism}

The hyperspherical harmonics (HH) method was first introduced for the description of charged excitons in quantum wells~\cite{Ruan1999,Ruan2000JPCM} and subsequently applied to trions in quantum dots using either the standard HH expansion~\cite{Kezerashvili2008FBS} or correlated HH expansions~\cite{Xie2000EPJB,Xie2001PSSB}, the latter being employed to accelerate convergence in Coulomb-interacting three-particle systems in quantum dots. A comprehensive exposition of the HH method is provided in Refs.~\cite{JibutiShitikova1993,Avery1989,AveryAvery2018}.

The SVM with explicitly correlated Gaussian basis functions has been employed to obtain highly accurate few-body solutions for excitonic complexes such as excitons, trions, and biexcitons confined in semiconductor quantum dots within the effective-mass approximation. Complementary to this approach, trion calculations in quantum dots using the HH method are typically formulated as a three-dimensional three-body problem with external parabolic confinement \cite{Bao2003}. In this formulation, the HH expansion is performed in the six-dimensional configuration space, and the Coulomb interaction retains its three-dimensional form, while quasi-two-dimensional effects are incorporated through anisotropic confinement parameters. In Refs.~\cite{Kezerashvili2008FBS,KezerashviliTsiklauri2013}, the HH method was extended to describe two-dimensional trion, biexciton, and triexciton states confined in quantum dots under a magnetic field.

By contrast, the HH approach developed and applied for the first time in Refs.~\cite{KezerashviliTsiklauri2017FBS,Filikhin2018Nano,Kezerashvili2024PRB,Kezerashvili2024arXivXeneTrions,Kezerashvili2024PRB_TMDCHH} treats trions in atomically thin two-dimensional materials as intrinsically 2D three-body systems. In this framework, the relative motion is described in a four-dimensional hyperspherical space, and the interparticle interaction is governed by the Rytova–Keldysh potential, which accounts for nonlocal dielectric screening in layered materials. As a result, the corresponding hyperradial equations differ qualitatively from those obtained in conventional three-dimensional formulations.

To solve the three-body Schr\"odinger equation~(\ref{eq:RelativeSchrodinger}) for a trion in a two-dimensional configuration space, the authors employ the method of hyperspherical harmonics. Within this framework, the relative motion of three particles interacting via centrally symmetric pairwise potentials, either the Rytova--Keldysh potential~(\ref{RKeldysh}) or the modified Kratzer potential~(\ref{Kr25}), is mapped onto the effective motion of a single particle in a four-dimensional (4D) hyperspace with a reduced mass $\mu$. Following Refs.~\cite{Kezerashvili2024PRB_TMDCHH,Kezerashvili2024PRB,Kezerashvili2024arXivXeneTrions}, the solution of Eq.~(\ref{eq:RelativeSchrodinger}) is obtained by introducing hyperspherical coordinates in the four-dimensional configuration space. The hyperradius is defined as
\begin{equation}
\rho = \sqrt{x_i^2 + y_i^2},
\end{equation}
while the hyperangular variables are collected into $\Omega_i = (\alpha_i,\varphi_{x_i},\varphi_{y_i})$, with $x_i = \rho \cos\alpha_i$ and $y_i = \rho \sin\alpha_i$.

In hyperspherical coordinates, the Schr\"odinger equation takes the form
\begin{equation}
\left[
-\frac{\hbar^2}{2\mu}
\left(
\frac{\partial^2}{\partial \rho^2}
+ \frac{3}{\rho}\frac{\partial}{\partial \rho}
- \frac{\hat{K}^2(\Omega_i)}{\rho^2}
\right)
+ \sum_{j=1}^{3} V(\rho \cos\alpha_j)
- \mathcal{E}
\right]
\Psi(\rho,\Omega_i)
= 0,
\label{eq:schrodinger_hyper}
\end{equation}
where $\hat{K}^2(\Omega_i)$ is the grand angular momentum operator in four dimensions \cite{JibutiShitikova1993,Avery1989,AveryAvery2018}.

The wave function is expanded in the complete orthonormal basis of hyperspherical harmonics,
\begin{equation}
\Psi(\rho,\Omega_i)
=
\rho^{-3/2}
\sum_{K\lambda}
u_{K\lambda}(\rho)
\,
\mathcal{Y}^{L}_{K\lambda}(\Omega_i),
\label{eq:HH_expansion}
\end{equation}
where $K$ is the grand angular momentum, $\lambda$ denotes the set of partial angular momenta, and $\mathcal{Y}^{L}_{K\lambda}$ are eigenfunctions of $\hat{K}^2$: $\hat{K}^2\mathcal{Y}^{L}_{K\lambda}=K(K+2)\mathcal{Y}^{L}_{K\lambda}$ \cite{Avery1989,AveryAvery2018}.

Substituting Eq.~(\ref{eq:HH_expansion}) into Eq.~(\ref{eq:schrodinger_hyper}) and integrating over the hyperangular variables yields a system of coupled differential equations for the hyperradial functions,
\begin{equation}
\left[
\frac{d^2}{d\rho^2}
-
\frac{(K+1)^2 - \tfrac{1}{4}}{\rho^2}
+ \kappa^2
\right]
u_{K\lambda}(\rho)
=
\frac{2\mu}{\hbar^2}
\sum_{K'\lambda'}
W_{K\lambda,K'\lambda'}(\rho)
\,u_{K'\lambda'}(\rho),
\label{eq:coupled_radial}
\end{equation}
where $\kappa^2 = 2\mu B/\hbar^2$, with $B$ being the trion binding energy.

The effective coupling potential is given by
\begin{equation}
W_{K\lambda,K'\lambda'}(\rho)
=
\int
\mathcal{Y}^{L*}_{K\lambda}(\Omega_i)
\left[
\sum_{j=1}^{3} V(\rho \cos\alpha_j)
\right]
\mathcal{Y}^{L}_{K'\lambda'}(\Omega_i)
\, d\Omega_i .
\label{eq:effective_potential}
\end{equation}

The coupled system of equations~(\ref{eq:coupled_radial}) thus describes the motion of an effective particle in 4D configuration hyperspace in a centrally symmetric coupling field defined by Eq.~(\ref{eq:effective_potential}).

To summarize, the HH method reformulates the three-body trion problem—two identical charge carriers and a distinguishable third—in terms of mass-scaled Jacobi coordinates and a single hyperradial coordinate, thereby mapping the two-dimensional three-body Schrödinger equation onto a coupled-channel problem in a four-dimensional hyperspace. The total wave function is expanded in a complete HH basis over the hyperangles, leading to a system of coupled hyperradial equations whose controlled truncation provides a systematic, nonperturbative treatment of three-body correlations. This approach proved effective for $X^-$ and $X^+$ trions in 2D materials, where the HH expansion captures correlation and symmetry effects beyond simple variational ansätze. In modern atomically thin semiconductors, the HH formalism has been adapted to include physically realistic screened interactions, such as the Rytova–Keldysh potential, and has been successfully applied to compute trion binding energies and systematic trends across TMDC \cite{KezerashviliTsiklauri2017FBS,Filikhin2018Nano,Kezerashvili2024PRB_TMDCHH} and buckled Xenes, including their tunability under external electric field \cite{Kezerashvili2024PRB,Kezerashvili2024arXivXeneTrions}.

Although excitonic systems are inherently many-body in nature, they can be
effectively treated within few-body frameworks
\cite{Kezerashvili2019}. The
Faddeev equations and the hyperspherical harmonics method, originally
developed in nuclear and atomic physics, have been adapted to
two-dimensional three-body systems calculations in
Refs.~\cite{Kezerashvili2008FBS,Braun1,Braun2,Mohseni2023PRB}.

\section{Benchmarking and Spectral Complexity of Trions in 2D Monolayers
}
\label{Sect_7}

Trions play a central role in the valley dynamics of emerging
2D semiconductors. In monolayers, trion properties are governed by coupled spin and valley degrees of freedom. The absence of inversion symmetry, together with strong spin--orbit coupling, leads to a large
valence-band splitting in monolayer TMDCs \cite{Xiao2012PRL}. This splitting
removes the spin degeneracy of electron and hole states in the $K$ and
$K^{\prime}$ valleys, allowing excitation of carriers with distinct
spin--valley configurations. The valley degree of freedom has been
extensively investigated in molybdenum- and tungsten-based dichalcogenide
monolayers using helicity-resolved photoluminescence, time-resolved
photoluminescence, reflectance spectroscopy, and the time-resolved Kerr
rotation technique \cite{Mak2012,Li2,WSe2
Jones,Zeng2012,Sallen2012,bZu,WSe2Wang,Wang2015,MacNeill2015,Yan2016,Gao2016,Huang2017,Hao2017}.

The absolute minima of the conduction band and the absolute maxima of the
valence band in TMDC monolayers are located at the two non-equivalent
hexagonal Brillouin-zone corners, $K$ and $K^{\prime}$. The lack of
inversion symmetry combined with spin--orbit interaction results in coupled
spin and valley degrees of freedom, such that the circular polarization of
absorbed or emitted photons is directly linked to a specific valley, $K$ or
$K^{\prime}$ \cite{Xiao2012PRL,Sallen2012}. These valley-contrasting optical
selection rules enable selective excitation and manipulation of the valley
index. Experimental studies \cite{Mak2012,Li2} demonstrate that each valley
can be selectively excited by light of a given helicity. With the
additional valley index $\tau$ alongside spin, one distinguishes between
two types of trions: intravalley and intervalley. For intravalley trions $\tau _{1}=\tau _{2}$, while for
intervalley trions $\tau _{1}\neq \tau _{2}$.

\begin{figure}[h!]
\centering
\includegraphics[width=13.0cm]{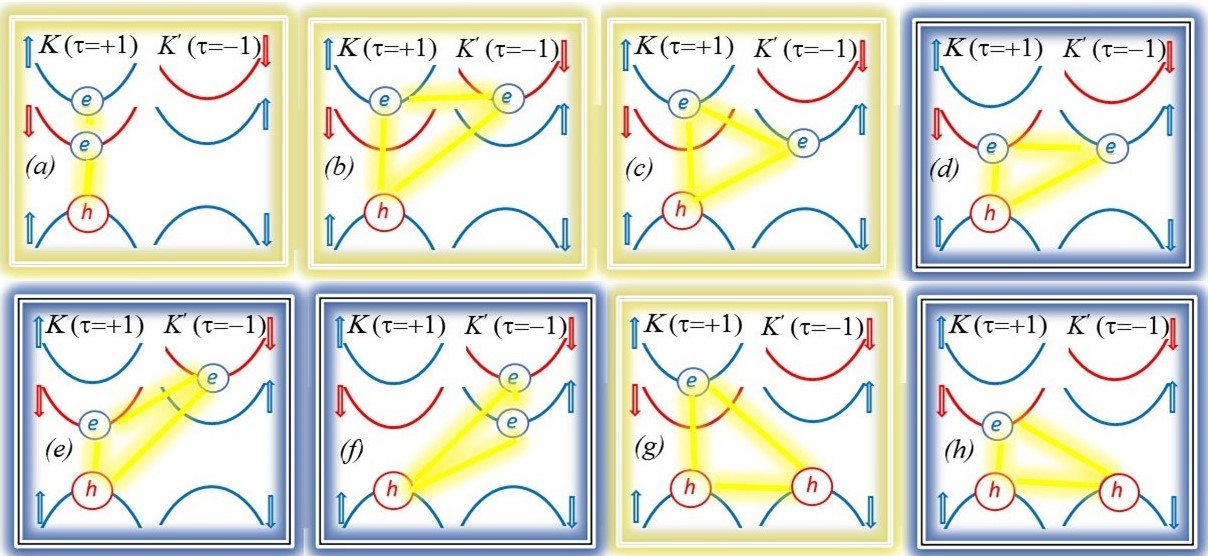}
\par
\caption{(Color online) Schematic illustration of WSe$_{2}$ low-energy band
structure and the spin-valley configurations of the constituent charge
carriers. The topmost spin-subband for the valence band and the lower and
upper spin-orbit split conduction band are shown. Arrows denote bands
with up (down) spin. The hole has the opposite spin of the valence electron
\protect\cite{Robert2017}. Light and dark rectangles indicate the bright and
dark trions, respectively. $(a)$, $(b)$, $(c)$, $(d)$, and $(e)$ correspond
to $X^{-}$ trions. $(g)$ and $(h)$ correspond to $X^{+}$ trions. The bright
trions emit circularly polarized light in the out-of-monolayer plane
direction, while the dark trions emit vertically polarized light in the
in-monolayer plane direction \protect\cite{Liu2019}. Lines indicate
interaction between three charged particles. The remaining configurations
are the time reversal of those shown in the figure. Adapted from Ref. \cite{Kezerashvili2024PRB_TMDCHH}.}
\label{TrionsValle}
\end{figure}

A symmetry analysis of the
different $X^{-}$ and $X^{+}$ trion states, classified by the valley
configurations as well as the spin configuration, is performed \cite%
{Ganchev,CourtadeSemina,NScRev2015,Fu2019,Kezerashvili2024PRB_TMDCHH}. Depending on the spin
configuration of the electron--hole pairs, intravalley excitons of TMDC
monolayers can be either optically bright or dark. These analyses \cite%
{CourtadeSemina,NScRev2015,Hao2017,Fu2019,Kezerashvili2024PRB_TMDCHH} show that the negative and
positive trions can be optically bright or dark, as illustrated in Fig. \ref{TrionsValle}. Experimental studies \cite%
{DarkT2017,Liu2019} have revealed evidence of dark trions in monolayer WSe$%
_{2}$ under a strong in-plane magnetic field.

The structural and spin-valley complexity of TMDC monolayers leads to rich
energetic properties \cite{light-valley,substrate,valley,CourtadeSemina,
revealing}. The $K$ and $K^{\prime}$ valleys, separated in momentum space,
together with strong spin-orbit coupling, give rise to singlet and triplet
trion configurations \cite{valley}.
A hybrid parameter-free theory, combining configuration interaction and Green’s function approaches, was proposed to study charged excitations \cite{Torche2019}. This fully first-principles many-body framework, based on the GW–Bethe–Salpeter approach, allows for the investigation of neutral excitons and charged trions in monolayer TMDCs. The authors computed trion fine-structure splittings arising from exchange interactions and spin–valley coupling, providing quantitative predictions for singlet–triplet energy separations and clarifying the microscopic origin of trion fine structure beyond traditional effective-mass models.

Negative trions can be optically bright or dark depending on the spin configuration of their constituent electron--hole pair: radiative recombination occurs when the electron and hole spins are aligned, whereas spin mismatch---typically involving the lower spin--orbit--split conduction band---suppresses emission, leading to dark trions; such dark trion states were theoretically predicted in Refs.~\cite{Deilmann2017,Danovich2017}. The intravalley $X^{-}$ trion exists in a spin-singlet state, whereas the intervalley $X^{-}$ trion can form spin-singlet or spin-triplet states. Consequently, these excitonic complexes can manifest as either spin-forbidden or momentum-forbidden dark states. An intriguing feature is that in cross-circularly polarized experiments, trions created in the $K$ valley in the singlet state (Fig.~\ref{TrionsValle}$a$) are converted to intervalley singlet trions with an electron in the $K^{\prime}$ valley (Fig.~\ref{TrionsValle}$c$) via spin flip and electron--hole exchange interaction \cite{Singh2016}.

In contrast to $X^{-}$, the positive trion $X^{+}$ consists of two holes occupying the topmost valence subbands in the $K$ and $K^{\prime}$ valleys and an unpaired electron in one of the conduction subbands; by the Pauli exclusion principle, it can exist only in the intervalley configuration. The $X^{+}$ trion is optically bright when the electron and hole spins are aligned (Fig.~\ref{TrionsValle}$g$) and dark otherwise (Fig.~\ref{TrionsValle}$h$).

Now let's start the comparison of the results obtained in different theoretical and computational approaches.
The theoretical investigation of trion binding energies in monolayer TMDCs reveals a dependence on material properties, the input parameters such as electron and hole effective mass, the screening distance chosen within the computational methodology, and the underlying physical approximations. Table \ref{SumupMet} highlights a general consensus across diverse theoretical frameworks regarding the magnitude of these energies, typically falling within the range of 20 to 35 meV for common TMDCs such as MoS$_2$, MoSe$_2$, WS$_2$, and WSe$_2$. This overall agreement suggests that, despite methodological differences, the effective-mass framework combined with screened Coulomb interactions provides a reliable description of trion binding in these systems.

\begin{table}[t]
\centering
\caption{Comparison of $X^{-}$ trion binding energies in representative TMDC monolayers. Results are compiled from various theoretical frameworks, each utilizing specific sets of electron and hole effective masses and dielectric screening lengths as input parameters. Values are provided in meV.
}
\begin{tabular}{lccccc}
\toprule
Method & MoS$_{2}$ & MoSe$_{2}$ & WS$_{2}$ & WSe$_{2}$ & Reference \\
\midrule
\multicolumn{6}{l}{\textit{Deterministic Few-Body Approach}} \\
VM   & 26    & 21    & 26    & 22    & \cite{Berkelbach2013} \\
    & 31.6  & 27.7  & 32.4  & 28.3  & \cite{Chang2021Trion} \\
SVM  & 33.7  & 28.2  & 33.8  & 29.5  & \cite{Kidd2016} \\
ED   & 31.82 & 27.84 & 32.60 & 28.55 & \cite{Kumar2025} \\
     & 31.7 & 27.7 & 34.2 & 28.4 & \cite{Fey2020} \\
GEM     & 33.7 & 28.2 & 33.8 & 29.5 & \cite{Tenorio2026GEM2D} \\
\midrule
\multicolumn{6}{l}{\textit{Quantum Monte Carlo Approach}} \\
DMC  & 33.8  & 28.4  & 34.0  & 29.5  & \cite{Mayers2015} \\
PIMC & 32.0   & 27.7   & 33.1   & 28.5   & \cite{Kylanpaa2015} \\
\midrule
\multicolumn{6}{l}{\textit{Faddeev Equations}} \\
FE   & 49.6  & ---   & ---   & ---   & \cite{Mohseni2023PRB} \\
\midrule
\multicolumn{6}{l}{\textit{Hyperspherical Harmonics}} \\
HH   & 32.8  & 27.6  & 33.1  & 28.3  & \cite{{Filikhin2018Nano},Kezerashvili2024PRB_TMDCHH} \\
\bottomrule
\end{tabular}
\label{SumupMet}
\end{table}
Deterministic few-body approaches, such as the variational method and exact diagonalization, provide foundational estimates. Early variational work by Berkelbach et al.~\cite{Berkelbach2013}  established binding energies near 26 meV for WS$_2$, while more recent high-precision ED calculations \cite{Kumar2025} report values around 32.6 meV. The SVM \cite{Kidd2016}, known for its high accuracy in few-body problems, yields slightly higher values, such as 33.8 meV for WS$_2$. The GEM \cite{Tenorio2026GEM2D} shows excellent agreement with the SVM \cite{Kidd2016}. This progression in values highlights how refined computational treatments of electron-hole correlations and non-local screening can lead to more robust predictions of trion binding.

Both DMC and PIMC provide numerically exact benchmarks that do not rely on a specific wavefunction ansatz. DMC results show a high degree of agreement with SVM. PIMC values are lower, illustrating the subtle differences in how stochastic methods handle the complex many-body interactions in these 2D systems.

The use of Faddeev equations and the hyperspherical harmonics method provides a rigorous treatment of the three-body problem, approaches that are well established in nuclear physics. While HH results 
align well with other methods, the Faddeev approach has produced significantly higher outliers, such as 49.6 meV, suggesting a higher sensitivity to the choice of potential and basis set in that specific framework.

An analysis of the material-specific trends summarized in Table~\ref{SumupMet} consistently indicates that sulfur-based TMDCs exhibit higher trion binding energies than their selenium-based counterparts. For instance, across nearly all methods, the binding energy for MoS$_2$ is roughly 4-5 meV greater than for MoSe$_2$. This trend is primarily attributed to the smaller effective masses and different dielectric screening properties associated with selenium, which reduce the overall Coulombic attraction between the constituent carriers.

To conclude, the robustness of these binding energies—which are significantly larger than those found in traditional 3D semiconductors—is a direct consequence of the reduced dimensionality and non-local dielectric screening characteristic of 2D materials. The strong agreement between independent methods like DMC, SVM, ED, GEM, and HH reinforces the validity of the effective-mass model when combined with an appropriate interaction potential, such as the Rytova-Keldysh potential, to describe charged excitonic complexes in the monolayer limit.

Many studies have been devoted to calculating the binding energies and wave
functions of excitonic complexes in monolayer TMDCs (see reviews
\cite{NScRev2015,BekReichman,Durnev2018,Kezerashvili2019,Suris2022}).
Theoretical investigations of negatively and positively charged trions,
$X^{-}$ and $X^{+}$, have employed a broad spectrum of analytical and
numerical techniques yielding binding energies in good agreement with experimental observations.

Beyond the deterministic \cite{Berkelbach2013,Chang2021Trion,Kidd2016,Kumar2025,Fey2020,Tenorio2026GEM2D,VargaNano2015,Varga2020} and quantum Monte Carlo \cite{Mayers2015,Velizhanin2015PIMC,Kylanpaa2015,Szyniszewski2017,Mostaani2017,Marsusi2022,Mostaani2023} approaches, extensive theoretical studies have utilized complementary techniques. These include effective-mass three-body models with nonlocal dielectric screening, first-principles many-body approaches (GW/BSE and configuration interaction), time-dependent density-matrix functional theory, and exciton–polaron formalisms that incorporate many-body interactions with the Fermi sea, and others.

The stochastic variational method with correlated Gaussian bases
\cite{VargaNano2015,Kidd2016,Varga2020} has been widely applied to investigate
trion binding energies and their dependence on screening length and
effective mass ratios. Extensions of this approach include the influence of
external magnetic fields and phonon recombination processes \cite{magnetic},
as well as the study of narrow resonance states using complex scaling and
stabilization techniques.

In Ref. \cite{Efimkin2017} authors developed a many-body theory for 2D semiconductors that describes trion absorption features as Fermi polarons, treating the trion not as an isolated three-body bound state but as a collective interaction between an exciton and the surrounding Fermi sea of electrons.

The theoretical description of trionic excitations has evolved from first-principles density frameworks to collective many-body models. In Ref.  \cite{TimeDepdensity_matrix_functional_theory} the authors utilized time-dependent density-matrix functional theory (TDDMFT) to treat trions as discrete three-body excitations. By incorporating a three-body kernel into the density-matrix equations of motion, this approach provides a rigorous microscopic derivation of the trion's internal structure and its role in the nonlinear optical response of monolayers like $\text{MoS}_2$.

In contrast, Efimkin and MacDonald \cite{Efimkin2017} introduced a many-body framework that reinterprets trion absorption features as Fermi polarons. In this picture, the trion is not viewed as an isolated bound state, but as an exciton "dressed" by its interactions with the surrounding resident carrier sea, described via a $T$-matrix scattering formalism. While the TDDMFT approach captures the specific quantum-mechanical bonds of the three-particle complex, the polaron model provides a more accurate description of the asymmetric spectral line shapes and the continuous evolution of quasiparticle energy levels as a function of carrier density.

Within the effective-mass framework, a microscopic description of trions in
monolayer TMDCs was developed in Ref.~\cite{Berkelbach2013}, where the
calculated binding energies were found to be in good agreement with
many-body Bethe--Salpeter equation results. Building on this foundation,
subsequent study introduced variational approaches based on
two-dimensional Slater-type orbitals, which enabled
efficient and accurate calculations of trion binding energies and
wave functions~\cite{Chang2021Trion}.

Alternative formulations include mapping the three-body problem with
logarithmic interactions to an effective one-particle problem treated with
the boundary-matching-matrix method \cite{Ganchev}. Trion states have also
been computed via direct diagonalization of the three-particle Hamiltonian
within the Tamm--Dancoff approximation
\cite{ZhumagulovTamm,ZhumagulovTamm2}. Both multi-band and single-band
models have been explored using finite-element and stochastic variational
methods \cite{model,Kidd2016}. The electronic and structural properties of excitons and trions in monolayer transition metal dichalcogenides were investigated within both multiband and single-band models \cite{model}. The energy eigenvalue equation was solved using the finite element method (FEM) as well as the SVM. While good agreement was found between the two approaches—and with other theoretical works—for excitons, the consistency for trions was less satisfactory.  This discrepancy stems from the neglect of angular correlations in the FEM treatment, which are vital for describing complex three-body states.

Substrate screening, doping, and
temperature effects further modify trion binding energies and fine-structure
splitting \cite{substrate,excited-state,Fu2019}. Optical spectroscopy,
including photoluminescence and reflectivity measurements, has played a
central role in extracting trion binding energies and revealing differences
between positive and negative trions
\cite{CourtadeSemina,substrate,coupling,Florian2018,revealing}.
Short-range Coulomb interactions and exchange effects have been proposed as
mechanisms for fine-structure splitting in WSe$_2$ monolayers
\cite{Glazov2020,coupling,CourtadeSemina}. Substrate-induced screening and
polarization are also known to reduce binding energies in supported
structures \cite{substrate}.

The many-body nature of excitons and trions in doped 2D semiconductors, particularly TMDCs, has been
extensively investigated in recent literature \cite{Chang2018, Chang2019, Rana2020}. These studies highlight a fundamental crossover from isolated few-body trion states at low doping to an exciton-polaron regime at higher carrier densities. Using a composite-boson formalism, Ref.~\cite{Chang2018} demonstrated that excitons become dressed by Fermi sea excitations, with energy features shifting due to many-body screening and exchange. This was expanded by Chang and Reichman \cite{Chang2019} via a diagrammatic $T$-matrix approach, which mapped doping-induced spectral shifts and broadening onto polaronic quasiparticle dynamics. Further, Rana et al. \cite{Rana2020} employed linear-response theory and a coupled Schrödinger-like framework to model optical conductivity, showing how strong Coulomb correlations hybridize these states. Collectively, these studies establish that a purely few-body description is
insufficient in doped 2D systems. Instead, a many-body framework is required
to capture the continuous evolution from isolated trionic bound states to
many-body dressed quasiparticle, thereby providing a unified theoretical basis
for interpreting doping-dependent optical experiments in TMDC monolayers.

More recently, high-lying excitons and trions involving carriers with negative effective mass were investigated within a variational framework
\cite{Glazov2023}, demonstrating the crucial role of nonparabolic conduction
band dispersion in stabilizing such bound states.

Studies on the temperature dependence of excitonic features in TMDCs reveal the critical role of phonon scattering and thermal distribution in quasiparticle stability. In Ref. \cite{Christopher2017SciRep} authors utilized temperature-dependent photoluminescence to demonstrate that in monolayer MoS$_2$, the trion-to-exciton intensity ratio decreases with rising temperature due to the thermal dissociation of the three-body complex. More recently, Nguyen et al. (2024) \cite{Nguyen2024SciRep} extended this analysis to WSe$_2$, identifying that the redshift and broadening of both exciton and trion peaks follow the O'Donnell-Chen model \cite{ODonnell1991}, while emphasizing that the trion binding energy remains relatively robust against thermal fluctuations compared to the significant changes in non-radiative recombination rates.

\begin{table}[t]
\centering
\caption{Trion binding energies for representative 2D semiconductors, including TMDC monolayers and phosphorene, under different dielectric environments: suspended, on a hBN substrate, and encapsulated by hBN. Energies are given in meV.}
\begin{tabular}{lccc}
\toprule
Material & Suspended & On substrate & Encapsulated \\
\midrule
MoS$_2$  & 32.12 & 21.59 & 15.83 \\
MoSe$_2$ & 32.00 & 22.60 & 17.18 \\
WS$_2$   & 34.49 & 23.60 & 15.94 \\
WSe$_2$  & 33.15 & 21.93 & 16.14 \\
Phosphorene  & 51.77 & 32.65 & 22.82 \\
\bottomrule
\end{tabular}
\label{TMDCvsMedia}
\end{table}

Interestingly enough, in \cite{photolum}, authors demonstrated that trion-like features observed in the photoluminescence of monolayer TMDCs can originate from virtual trion states formed through many-body exciton–carrier scattering processes, where Coulomb interactions with the Fermi sea dynamically dress the exciton and enable recombination through intermediate, non–fully bound three-body configurations rather than through stable trion states.

In the work by Cavalcante \textit{et al.}~\cite{Cavalcante2018}, trion binding energies were theoretically studied for several two-dimensional materials, with particular emphasis on TMDCs and black phosphorus under an external electric field. The binding energies calculated in the absence of an electric field are summarized in Table~\ref{TMDCvsMedia}. The data clearly indicate that trion binding energies are strongly influenced by both intrinsic material parameters and the surrounding dielectric environment. Comparing different materials, phosphorene monolayer exhibits a significantly stronger binding energy when suspended than TMDC family, which ranges between approximately 32 and 35 meV. Moreover, the comparatively larger binding energies in phosphorene highlight the role of its strongly anisotropic effective masses and reduced dielectric screening, which significantly enhance Coulomb correlations relative to isotropic TMDC systems. Beyond material type, the environmental conditions impose a drastic influence on stability; as the dielectric screening increases from a suspended state to a substrate-supported and eventually a fully encapsulated configuration, the binding energies decrease monotonically. For all listed semiconductors, encapsulation in hBN approximately halves the binding energy relative to the suspended case—for example, the binding energy of WS$_2$ drops from 34.49 meV to 15.94 meV. This highlights that while trions are robustly bound in 2D systems, their precise energy levels and stability are highly tunable through dielectric engineering of the surrounding medium.

In Ref. \cite{Pucher2024AFM}, the electrostatic control of photoluminescence of monolayer MoS$_2$ at room temperature via integration of free-standing BaTiO$_3$ (BTO), a ferroelectric perovskite oxide is studied. The authors demonstrate that trion properties are highly sensitive to the dielectric environment: the ultrahigh-$\kappa$ screening combined with polarization-induced doping enables a large tunability of the trion binding energy from $\sim 40$ to 100 meV at room temperature. A comparison with hBN shows that the ultrahigh-$\kappa$ BTO dielectric provides significantly stronger electrostatic control, leading to enhanced tunability of trion properties, reduced operating voltages, and thus lower power consumption.

Although early studies of trions in phosphorene predicted extraordinarily large trion binding energies exceeding 100 meV based on effective-mass models \cite{Chaves2016PRB}, subsequent ab initio studies utilizing the Bethe-Salpeter Equation refined these values to approximately 70--90 meV \cite{DeilmannThygesen2018}.
Recent theoretical studies have provided a more refined understanding
of trions in phosphorene, emphasizing the crucial role of anisotropy,
dielectric screening, and many-body correlations. In Refs. ~\cite{Liang2024PCCP,Sheng2024APhA_MB} demonstrated that trion binding
energies in phosphorene are highly sensitive to the dielectric
environment, with substrate-induced screening substantially reducing
the effective binding energy compared to earlier estimates,
yielding values on the order of $\sim 70$ meV in screened few-layer
configurations. In a complementary many-body treatment, in Ref.~\cite{Sheng2024APhA_MB} the authors developed a theoretical framework for
trions in two-dimensional nanostructures and applied it to phosphorene
nanoflakes, showing that realistic dielectric modeling reconciles
theoretical predictions with experimental observations.
Furthermore, Zhong \textit{et al.}~\cite{Zhong2024PCCP} investigated
electric-field effects in  quasi-1D phosphorene
nanostructures, demonstrating that external fields can enhance
excitonic binding by increasing carrier localization, thereby offering
an additional mechanism for tuning Coulomb correlations in anisotropic
phosphorene systems.

 Thus, advanced many-body calculations clarified the role of dielectric screening and band anisotropy, showing that the apparent magnitude of trion binding energies in monolayer black phosphorus depends sensitively on the reference energy and environmental screening \cite{Chaves2016PRB,DeilmannThygesen2018,Liang2024PCCP,Sheng2024APhA_MB}. Together, these works establish phosphorene as a viable platform for stable trions, while highlighting the importance of anisotropy and dimensional crossover effects in determining their binding and optical properties. Transition metal trichalcogenides also represent quasi-one-dimensional van der Waals semiconductors with strong Coulomb interactions and reduced dielectric screening, which can support strongly bound excitonic complexes, including trions \cite{Chen2023TMTCReview}.

\section{Trions in External Electric and Magnetic fields}
\label{Sect_8}

The influence of external electric fields on trions in two-dimensional materials has attracted considerable attention due to the tunability of their optical response, internal structure, and binding energies. Existing studies can broadly be categorized into two main directions: (i) works that explicitly compute the electric-field dependence of trion binding energies within a microscopic three-body framework, and (ii) investigations that analyze field-induced shifts of optical transition energies (Stark effect) without directly evaluating modifications of the underlying binding energy.

The behavior of trions in 2D materials under external magnetic fields has revealed rich valley, spin, and many-body phenomena. Magnetic fields lift valley degeneracy, modify spin–valley coupling, and affect both optical transitions and the binding energies of charged excitonic complexes. These effects provide valuable insight into their internal structure and correlated dynamics of trions in reduced dimensionality.

\subsection{Trions in 2D Materials under External Electric Fields}

\begin{figure}[b]
\centering
\includegraphics[width=10cm]{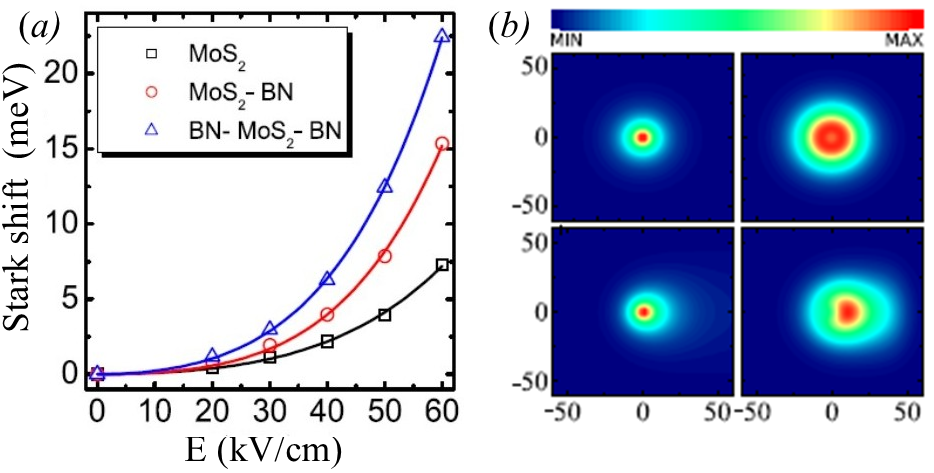}
\caption{($a$) Numerically obtained field dependence of
the trion binding energy in monolayer MoS$_2$ in the freestanding case, on a hBN substrate, and encapsulated by hBN. The values
of polarizability and hyperpolarizability (fitting parameters for the curves) are presented in Ref \cite{Cavalcante2018}. ($b$) Contour map of the square modulus of the trion wave function in freestanding MoS$_2$. The left
(right) column represents the exciton’s (trion’s) center of mass wave
function in the absence of the electric field in the first row, and for a
$E$ = 60 kV/cm field applied in the $x$ direction in the second row.
The color scale goes from 0 (blue, MIN) to 0.012 (red, MAX). Adapted from \cite{Cavalcante2018}.}
\label{fig:StarkEfeect}
\end{figure}
Electric fields provide a means of tuning on both the internal structure and optical signatures of trions in two-dimensional materials. A systematic and rigorous theoretical treatment of excitons and trions under an applied electric field was developed by Cavalcante \textit{et al.}~\cite{Cavalcante2018}.  To account for the influence of the electric field on trions, one must add to the trion Hamiltonian~(\ref{eq:TrionHamiltonian})
the interaction term -$\boldsymbol{d} \cdot \mathbf{E}$, where $\boldsymbol{d}$ is the dipole moment and $\mathbf{E}$ is the external electric field. Within the effective-mass approximation and a screened Coulomb interaction, the authors~\cite{Cavalcante2018} computed the Stark response of both neutral and charged excitonic complexes in TMDCs and phosphorene, and analyzed the stability of trions as a function of field strength. Their results demonstrate that electric fields modify the internal structure of the three-body wave function, leading to a reduction of the binding energy beyond a critical field.

More generally, the energy shift of an excitonic complex with $s$-state symmetry in an electric field $\mathbf{E}$ can be expanded as \cite{Cavalcante2018}
\begin{equation}
\mathcal{E}(E) =\mathcal{E}+
\frac{1}{2}\alpha E^2
+\frac{1}{24}\gamma E^4,
\end{equation}
where $\alpha$ and $\gamma$ are the polarizability and hyperpolarizability, respectively. In monolayer TMDCs, intralayer trions typically exhibit a quadratic Stark effect, whereas in bilayer or layer-hybridized systems a linear contribution may arise due to a finite dipole moment.

Figure \ref{fig:StarkEfeect}$a$ illustrates the dependence of the trion binding energy in MoS$_2$ on an applied in-plane electric field due to the Stark effect. As the electric field strength increases, the trion binding energy decreases. The trion, consisting of three Coulomb-correlated charge carriers confined to a 2D sheet, becomes polarized under the applied field. This leads to a spatial separation of the electron and hole probability densities, which modifies the internal structure of the three-body wave function and shifts the trion energy relative to its zero-field value.
Moreover, Fig. \ref{fig:StarkEfeect}$a$ demonstrates the robustness of the trion state: no dissociation occurs even for fields up to 60 kV/cm, as evidenced by the wave function projections in Fig. \ref{fig:StarkEfeect}$b$ for the suspended MoS$_2$ case. In the absence of an electric field (top row of Fig. \ref{fig:StarkEfeect}$b$), both the excitonic and trionic components of the wave function are circularly symmetric. The trionic component exhibits a minimum at $  r = 0  $ due to electron-electron repulsion. The applied in-plane electric field induces deformation of these distributions (bottom row of Fig. \ref{fig:StarkEfeect}$b$), but they remain concentrated around the origin, indicating the absence of trion dissociation.

Trion binding energies depend on the dielectric environment (suspended, on-substrate, or encapsulated), but they remain sufficiently large to ensure stability under relatively strong electric fields, as shown in Fig. \ref{fig:StarkEfeect}$a$ MoS$_2$ case. Trions in phosphorene 
are particularly robust against applied fields, owing to the material's anisotropic effective masses and reduced dielectric screening. In monolayer phosphorene, for fields applied along the zigzag direction, even strengths as high as 100 kV/cm produce negligible shifts in binding energy (on the order of $\sim$1 meV or less) and virtually no change in the wave function distribution \cite{Cavalcante2018}.

This robustness supports the potential for using in-plane electric fields to drive charged excitons across the material plane without dissociation, which could enable novel optoelectronic applications such as energy transfer via trion transport.

In parallel with theoretical studies, several experiments and theory–experiment collaborations focus on field-tunable trion transition energies. Electrically tunable trion resonances have been observed experimentally in several systems. For example, Ross \textit{et al.}~\cite{Ross2013ElectricalControl} demonstrated gate-controlled conversion between neutral and charged excitons in monolayer TMDCs. In Ref.~\cite{Liu2019DarkTrions}, the authors reported gate-tunable dark trions in monolayer WSe$_2$. More recently, Perea-Causin \textit{et al.}~\cite{PereaCausin2024LayerHybridizedTrions} showed that an out-of-plane electric field in WSe$_2$ bilayers enables electrical control of layer-hybridized trion states and induces pronounced Stark red-shifts in photoluminescence spectra.

These studies collectively demonstrate that electric fields provide a powerful external knob for engineering trion stability, fine structure, and optical response in 2D  materials.

Let us consider trions in Xenes under an applied perpendicular electric field. In Ref.~\cite{Kezerashvili2024PRB}, we demonstrated within the
HH approach that perpendicular electric
fields in Xenes act as a direct control parameter of the three-body
Hamiltonian itself. In buckled 2D materials, an electric field creates an asymmetry between the two sublattices. This modifies the band gap and, consequently, the effective masses and the strength of Coulomb interactions between the constituent particles of the trion. Unlike conventional Stark-effect studies,
where the field primarily induces polarization of the excitonic complex, in Xenes, the applied field simultaneously modifies the effective mass and dielectric screening, leading to a strong and
nonperturbative tuning of trion binding energies.
This behavior originates from a qualitatively different physical mechanism that operates in Xenes, where a perpendicular electric field $E_\perp$ does not merely induce a Stark polarization of
the trion wave function, but instead directly modifies the
underlying single-particle band structure due to the intrinsic
buckled lattice geometry. In these materials, the two
sublattices are vertically displaced by a distance $2l$, where $l$ is half the buckling constant,
so that the application of $E_\perp$ produces an electrostatic
potential difference between sublattices. As a result, the
field-dependent band gap takes the form \cite{Tabert2014}

\begin{equation}
\Delta_{\xi\sigma}(E_\perp)
=
\left| \Delta_{\mathrm{SO}} \, \xi \sigma
- e E_\perp l \right|,
\end{equation}
\noindent
where $\Delta_{\mathrm{SO}}$ is the intrinsic spin–orbit coupling,
$\xi=\pm1$ denotes the valley index, $\sigma=\pm1$ the spin index,
$e$ is the elementary charge, and $l$ is half the buckling height.

The effective mass therefore, becomes field dependent,

\begin{equation}
m^*(E_\perp)
=
\frac{\Delta_{\xi\sigma}(E_\perp)}{2 v_F^2},
\end{equation}
with $v_F$ the Fermi velocity. Since the trion binding energy
in a two-dimensional system approximately scales as

\begin{equation}
\mathcal{E}_b \sim \frac{\mu(E_\perp)}{\varepsilon^2(E_\perp)},
\end{equation}
where $\mu(E_\perp)$ is the reduced mass and
$\varepsilon(E_\perp)$ the effective dielectric screening,
the perpendicular electric field renormalizes the parameters
of the few-body Hamiltonian rather than simply perturbing
the three-body wave function.

\begin{figure}[b]
\centering
\includegraphics[width=11 cm]{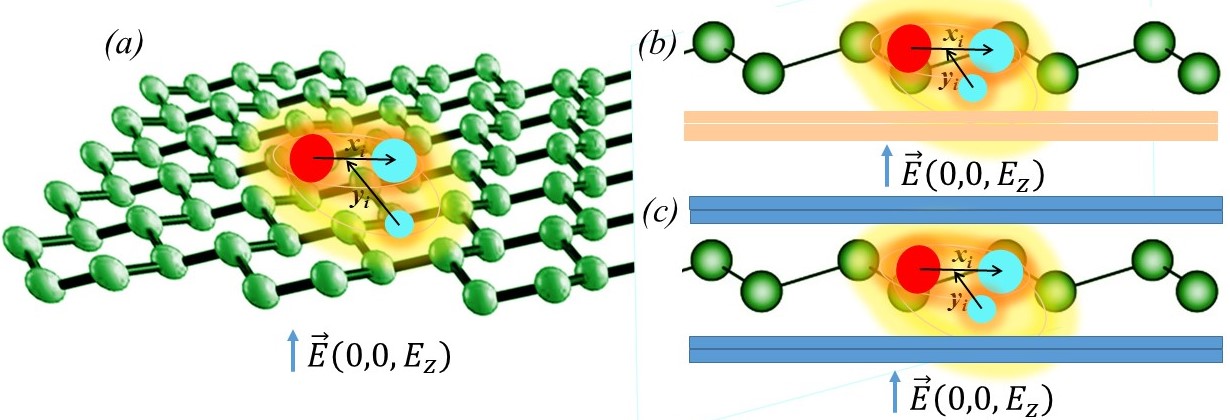}
\includegraphics[width=11.5 cm]{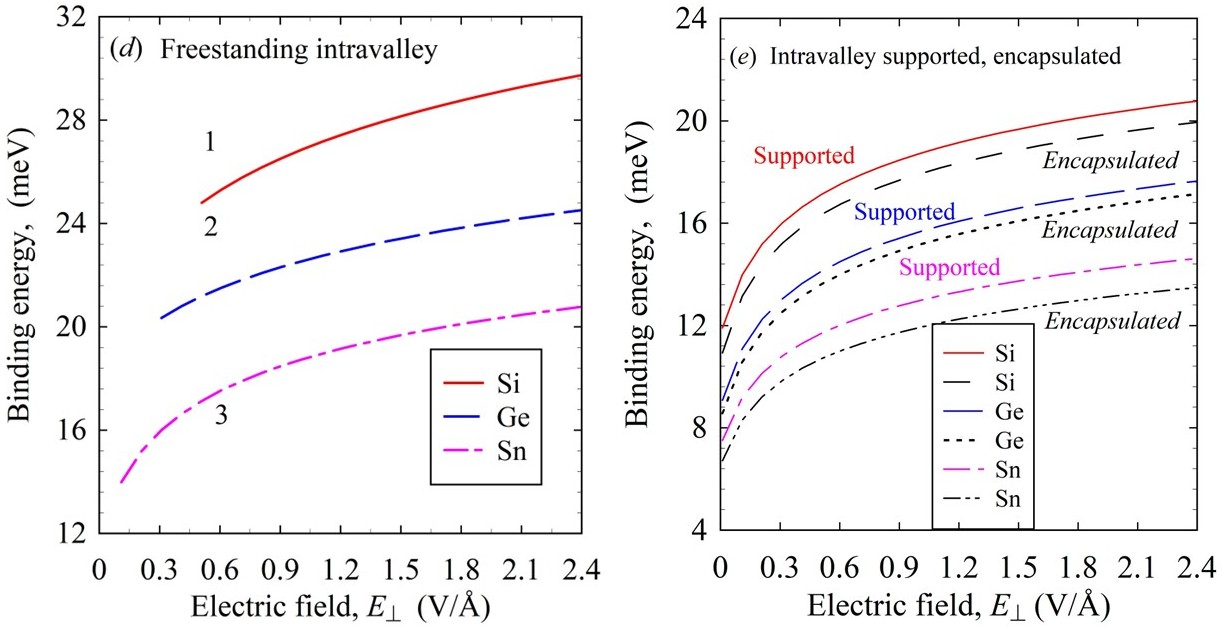}
\includegraphics[width=11 cm]{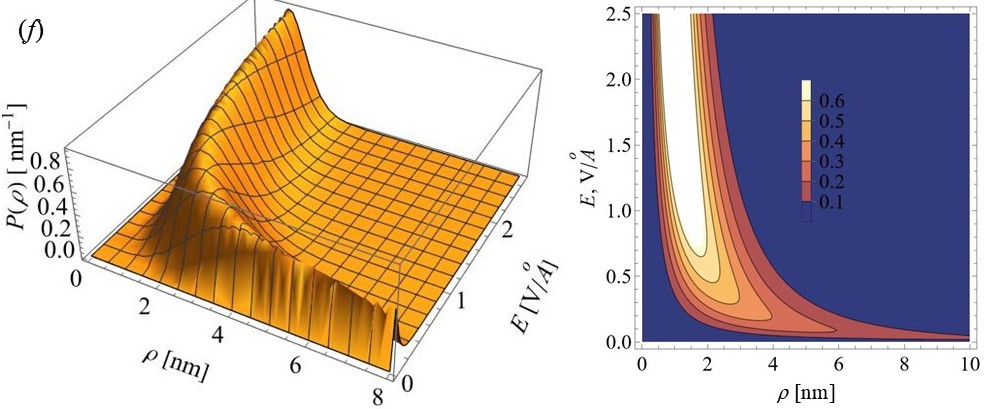}
\caption{Schematic representation of the X$^{-}$ trion in the external electric field perpendicular to the Xene layer in ($a$)
freestanding, ($b$) supported and ($c$) encapsulated Xenes monolayer. $x_{i}$
and $y_{i}$ are Jacobi coordinates for the partition $i$.
Dependencies of the binding energy of intravalley singlet state trions in silicene, germanene, and stanene on the applied electric filed. ($d$) Curves 1, 2, and 3 correspond to the $X^{-}$ trions formed by $A$ excitons coupling an electron. The curves for freestanding Xenes are truncated for the external electric field $E_{\perp}$ less than a critical field. ($e$) The curves correspond to $X^{-}$ trions in  Xenes monolayer supported on an SiO$_{2}$ substrate and hBN encapsulated. ($f$) Dependence of the probability distribution of three particles in freestanding silicene
on hyperradius $\rho$ and applied electric field for intravalley trion. Adapted from \cite{Kezerashvili2024PRB}.}
\label{Freestanding}
\end{figure}

This mechanism is fundamentally distinct from the conventional
Stark effect in TMDC monolayers, where the band structure is
essentially fixed and the electric field primarily induces
polarization of the correlated excitonic complex. In Xenes,
by contrast, $E_\perp$ tunes the band gap continuously and can
drive the system through regimes of enhanced or reduced
effective mass, thereby significantly strengthening or weakening
Coulomb correlations.

Consequently, the trion binding energy may exhibit a
non-monotonic dependence on the applied field. Near the
critical field at which the band gap closes
($eE_\perp l \approx \Delta_{\mathrm{SO}}$),
the effective mass becomes small and screening effects are
strongly modified, leading to substantial changes in the
stability of excitonic complexes. In certain regimes,
field-induced enhancement of binding is possible, while in
others, increased screening or mass reduction suppresses
the trion stability.

Thus, in Xenes the perpendicular electric field serves as a
direct control parameter of the few-body Hamiltonian itself.
This provides a unique platform for electrically tunable
many-body physics, enabling controlled manipulation of trion
binding energies, valley-spin selectivity, and potentially
field-driven excitonic phase transitions.

At nonzero electric fields, both the valence and conduction bands split into
upper subbands with a large gap (for $\xi = -\sigma$) and lower subbands with
a small gap (for $\xi = \sigma$). We refer to excitons formed from charge
carriers in the large-gap subbands as $A$ excitons, and those formed from
carriers in the small-gap subbands as $B$ excitons. The external electric
field affects the reduced masses of the $A$ and $B$ excitons differently,
primarily due to their distinct intrinsic band gaps. The direct bright $A$
and $B$ excitons consist of electrons and holes with parallel spins. In all
spin--valley configurations, the mass of the $A$ exciton exceeds that of the
$B$ exciton.

The intravalley $X^{-}$ and $X^{+}$ trions exist only in the spin-singlet state,
whereas intervalley trions can form either singlet or triplet states.
Figures~\ref{Freestanding}$d$ and \ref{Freestanding}$e$ present the binding
energies of intravalley trions formed by an $A$ exciton coupled to an electron
in freestanding (FS), SiO$_2$-supported, and hBN-encapsulated Xenes monolayers.
A comparison of the binding energies (BEs) shows that FS monolayers exhibit
significantly larger BEs due to the much weaker dielectric screening of the
Coulomb interaction. The binding energy decreases for Xenes supported on
SiO$_2$ and is smallest for hBN-encapsulated structures, demonstrating the
strong sensitivity of trion binding to the dielectric environment.

Intervalley singlet trions exhibit a qualitatively similar dependence on the
external electric field as intravalley trions. In contrast, intervalley
triplet trions in FS silicene and germanene monolayers have binding energies
approximately three times smaller than those of the corresponding singlet
states~\cite{Kezerashvili2024PRB}. Moreover, both intra- and intervalley trions
formed by a $B$ exciton coupled to an electron possess smaller binding
energies than those formed by an $A$ exciton, since in all spin--valley
configurations the mass of the $A$ exciton exceeds that of the $B$
exciton~\cite{Kezerashvili2024PRB}.

Figure~\ref{Freestanding}$f$ shows the probability distribution of the three
particles forming an intravalley trion in a silicene monolayer. The
corresponding distribution for intervalley trions differs slightly, with the
difference arising from the distinct three-particle effective masses of intra- and intervalley trions~\cite{Kezerashvili2024PRB}. An analysis of the probability
distribution as a function of the hyperradius $\rho$ and the external electric
field indicates that increasing the electric field enhances the trion binding
energy and leads to a more compact spatial structure. The stronger binding
therefore, increases the probability of trion formation.

In many Xenes configurations, the effective masses of electrons and holes are
not perfectly symmetric. As a result, $X^{+}$ and $X^{-}$ trions exhibit
distinct binding-energy profiles under the same external electric
field~\cite{Kezerashvili2024PRB}.

\subsection{Trions in 2D Materials under External Magnetic Field}

The existence of tightly bound trions in monolayer TMDCs was first established experimentally by Mak \textit{et al.}~\cite{Mak2013NatMat}, who demonstrated gate-controlled formation of charged excitons in monolayer MoS$_2$.
Subsequently, magneto-photoluminescence measurements have
demonstrated pronounced Zeeman splitting of trion resonances
in monolayer TMDCs. In a single issue of \textit{Nature Physics}, two seminal papers were published \cite{Srivastava2015, Aivazian2015NatPhys} that provided the first direct experimental evidence of how a magnetic field breaks the "valley degeneracy" in monolayer WSe$_2$. Together, these works established the definitive experimental foundation for the valley Zeeman effect in TMDCs. Aivazian \textit{et al.}~\cite{Aivazian2015NatPhys} demonstrated magnetic control of the valley pseudospin in WSe$_2$, highlighting the distinct magnetic responses of neutral and charged excitons.
Similarly, Mitioglu \textit{et al.}~\cite{Mitioglu2015NanoLett}
performed high-field optical studies of WSe$_2$, resolving
field-dependent exciton and trion transitions, further clarifying the diamagnetic
shifts and $g$-factors of these complexes.
Lyons \textit{et al.}~\cite{Lyons2019NatCommun} expanded this scope by
investigating the valley Zeeman effect in inter- and intravalley
trions, demonstrating different $g$-factors for various
charged excitonic species.
Magnetically tunable dark trions were identified by
Liu \textit{et al.}~\cite{Liu2019PRLdarktrions},
where the magnetic field enabled the separation of bright and dark
charged complexes via their characteristic Zeeman splittings. In Ref. \cite{Kuznetsov2018PRB} experimentally investigated three-particle electron–hole complexes in high-mobility 2D electron systems under strong magnetic fields, revealing the formation and optical signatures of bound and metastable charged exciton states. Their results clarified the interplay between Coulomb interactions, Landau-level structure, and many-body effects in magneto-trion spectra.  Kapu\'sci\'nski \textit{et al.}~\cite{Kapuscinski2020PCCP} experimentally studied singlet and triplet trions in a WS$_2$ monolayer under an external magnetic field, showing that the valley Zeeman effect modifies the energies and relative intensities of trion states in polarization-resolved optical spectra.

On the theoretical side, variational calculations of singlet and triplet states of $X^{-}$ and $X^{+}$ trions confined in semiconductor quantum wells under a perpendicular magnetic field within the effective-mass approximation were carried out in Refs.~\cite{PRB63115302,Riva2001PRB64}. These studies analyzed how the trion binding energies, energy spectra, and correlation properties depend on the quantum-well width and magnetic-field strength, predicting singlet–triplet transitions and demonstrating good agreement with experimental observations in GaAs/AlGaAs quantum wells. \textit{Ab-initio} many-body calculations \cite{Deilmann2017,Deilmann2020PRL} showed the role of dark excitons and trions, whose energies and optical activity can be modified by magnetic fields and spin–orbit coupling. Complementary to these approaches, Ref.~\cite{Sergeev2005M} developed a semi-analytical method providing universal estimates of trion binding energies. Further advancing these treatments, Ref.~\cite{Wojs2007PRBTrionED} performed detailed studies incorporating Landau-level (LL) structure and many-body correlations. Together, these works established the theoretical foundation of magnetotrion physics in conventional quasi-two-dimensional semiconductor systems.

The emergence of atomically thin TMDC monolayers introduced a new regime characterized by strong Coulomb interactions, reduced dielectric screening, and valley degrees of freedom. In contrast to GaAs-based structures, TMDC monolayers exhibit non-hydrogenic screening and pronounced spin–valley coupling, requiring modified theoretical frameworks. This shift has motivated renewed efforts to treat $c.m.$--internal coupling and Landau-level mixing in strictly two-dimensional materials.

In Ref.~\cite{magnetic}, the authors systematically investigated excitons, trions, and biexcitons in TMDC monolayers. Within an effective-mass framework, binding energies were computed using the stochastic variational method, demonstrating how Landau quantization and magnetic confinement modify the stability of these three-body complexes. More recently, Ref.~\cite{Chang2022JCP} addressed a central difficulty of the magnetotrion problem: the non-separability of $c.m.$ and internal degrees of freedom. By applying a unitary transformation to the three-particle Schr\"odinger equation, the translational--internal entanglement was explicitly accounted for. Finally, Ref.~\cite{Aleksandrov2024PRBTrionMagnetic} critically reassessed the two-dimensional trion problem, demonstrating that the exact inclusion of Coulomb-induced LL mixing is essential for reliable predictions of binding energies, which are otherwise significantly underestimated in high-field regimes.

The application of a perpendicular magnetic field introduces
additional confinement and modifies both the kinetic energy
and the internal structure of trions in 2D
semiconductors. In the presence of a uniform magnetic field
$\mathbf{B}=B\hat{\mathbf{z}}$, the canonical momentum is replaced
by the prescription
$\mathbf{p}_i \rightarrow
\mathbf{p}_i - \frac{q_i}{c}\mathbf{A}(\mathbf{r}_i)$,
where $q_i$ is the charge of the $i$th particle and
$\mathbf{A}$ is the vector potential, typically chosen in the
symmetric gauge
\begin{equation}
\mathbf{A}(\mathbf{r}) =
\frac{1}{2}\mathbf{B}\times\mathbf{r}.
\end{equation}
For a trion, $X^{\pm}$ in magnetic field, the Hamiltonian (\ref{eq:TrionHamiltonian}) becomes
\begin{equation}
\hat{H}_{m} =
\sum_{i=1}^{3}
\frac{1}{2 m_i}
\left[
\mathbf{p}_i
-
\frac{q_i}{c} \mathbf{A}(\mathbf{r}_i)
\right]^2
+
\sum_{i<j} V(r_{ij}),
\label{eq:Hmag}
\end{equation}
where $V(r_{ij})$ is the screened Coulomb interaction.

In contrast to the zero-field case, $B=0$, the $c.m.$ and relative motions no longer fully decouple due to the
magnetic vector potential. Therefore, the $c.m.$ motion cannot be
separated from the internal degrees of freedom in the conventional
way. As shown in Ref.~\cite{Dzyubenko2000}, the proper
theoretical treatment of trions in a magnetic field must be based on magnetic translation symmetry rather than
on naive separation of the c.m.\ motion. Because a trion carries
net charge, ordinary translational invariance is broken, standard
momentum conservation is replaced by magnetic translations, and
the components of the magnetic translation operator become
noncommuting.

The pseudomomentum operator was first introduced in the study of
excitons in a magnetic field by Gor'kov and Dzyaloshinskii
\cite{Gorkov1967}, where it was shown to commute with the
exciton Hamiltonian and is therefore conserved. Following
Refs.~\cite{Dzyubenko2000SSC,Dzyubenko2001}, for $X^\pm$ trions
in a uniform perpendicular magnetic field, the conserved magnetic-translation
(pseudomomentum) operator can be written as

\begin{equation}
\hat{\mathbf K}
=
\sum_{i=1}^{3}
\left[
\hat{\mathbf p}_i
-
\frac{q_i}{c}\mathbf A(\mathbf r_i)
+
\frac{q_i}{c}\,\mathbf B \times \mathbf r_i
\right],
\end{equation}
which satisfies
$[\hat{H}_{m},\hat{\mathbf K}]$ = 0.
However, for a charged trion with total charge
$Q=\sum_i q_i \neq 0$, the components of $\hat{\mathbf K}$ obey
the noncommutative algebra
\begin{equation}
[\hat K_x, \hat K_y]
=
- i\hbar\,\frac{Q B}{c},
\end{equation}
so that $\hat K_x$ and $\hat K_y$ do not commute, and so that $\hat K_x$ and $\hat K_y$ cannot be simultaneously
diagonalized.
Consequently, only $\hat{\mathbf K}^2$ is conserved and is used
to classify magnetotrion states. For $Q\neq 0$, this structure
leads to Landau-like quantization of the trion $c.m.$
motion and modifies the internal binding energy.

In Ref.~\cite{Chang2022JCP}, the trion is described by a three-particle Schr\"odinger equation with the magnetic field introduced through a vector potential in the symmetric gauge. 
Using coordinate and unitary transformations, the Hamiltonian is decomposed into: (i) a translational term describing Landau quantization of the $c.m.$ motion, (ii) an internal binding term, and (iii) a translational--internal coupling term, linear in the magnetic-field strength, accounting for $c.m.$--internal entanglement.

The eigenenergies and wavefunctions are obtained variationally, with the trial wavefunction constructed as a linear combination of products of internal trion states and Landau-level functions, explicitly incorporating $c.m.$--internal mixing. The internal part is expanded in Slater-type orbitals, and each eigenstate is characterized by a shell number, spin permutation number, total angular momentum, and degeneracy index. Applied to hole-doped WSe$_2$ monolayers up to 25~T, the method predicts excited trion states stabilized and bound by a perpendicular magnetic field, explaining bright and dark triplet structures in magneto-optical spectra.

A more rigorous treatment was presented in Ref.~\cite{Aleksandrov2024PRBTrionMagnetic}, where two-dimensional magnetotrions were reexamined. It was shown that neglecting Coulomb-induced mixing between LLs leads to substantial underestimation of binding energies in experimentally relevant fields, and that this deficiency persists even at very high fields.

By treating Coulomb mixing exactly and preserving translational symmetry via magnetic translations, the full trion Hamiltonian was numerically diagonalized. Additional discrete states were identified; for example, in GaAs a second triplet state appears near $B \approx 9\,\text{T}$. Exact LL mixing also shifts the continuum boundary and increases binding energies (e.g., in a GaAs quantum well at 30~T, from 1.298~meV to 4.680~meV).

A central conceptual result of ~\cite{Aleksandrov2024PRBTrionMagnetic} is the demonstration that
Coulomb-induced Landau-level mixing does not vanish even at high
magnetic fields. Although Landau-level spacing increases linearly
with $B$, the Coulomb matrix elements scale as $\sqrt{B}$, so
second-order corrections remain finite in the strong-field limit. Also, the work further provides a comprehensive database of matrix elements for computing magnetotrion spectra in a wide range of materials.

Collectively, both studies \cite{Chang2022JCP,Aleksandrov2024PRBTrionMagnetic} demonstrate that a reliable description of magnetotrions requires an explicit treatment of the coupling between $c.m.$ and internal degrees of freedom, and the exact inclusion of Coulomb-induced LL mixing, which are essential for reliable predictions of binding energies and discrete-state structure in quasi-two-dimensional systems.

Overall, these symmetry-consistent and Landau-level–complete approaches provide a rigorous framework for describing magnetotrions, highlighting the pivotal role of Coulomb-induced Landau-level mixing and $c.m.$–internal coupling in interpreting high-field experiments on 2D semiconductors.



\section{Concluding Remarks}
\label{Sect_9}

This review has traced the conceptual and methodological evolution of
trion physics from its origins in bulk semiconductors to its modern
realization in atomically thin materials. What began as an effective-mass
analogue of the negative hydrogen ion has developed into a rich and
quantitatively precise field of few-body physics in reduced
dimensionality. The dramatic enhancement of Coulomb interactions in
two-dimensional systems—driven by reduced dielectric screening and
quantum confinement—has elevated trions from weakly bound excitonic
complexes to robust quasiparticles with binding energies reaching
tens or even hundreds of meV.

A central theme emerging from this review is the complementarity
between few-body and many-body approaches. Deterministic variational
methods, stochastic variational techniques, hyperspherical harmonics,
Faddeev equations, and quantum Monte Carlo simulations now provide
benchmark-level predictions for binding energies and structural
properties of trions in two-dimensional semiconductors. At the same
time, experimental progress—particularly in TMDC monolayers and
anisotropic materials such as phosphorene—has confirmed the essential
role of dielectric environment, effective-mass anisotropy, and carrier
density in shaping trion spectra.

Several open directions define the outlook of the field.
First, a unified theoretical description that consistently bridges
few-body bound states and exciton–polaron regimes at finite doping
remains an active challenge. Second, the treatment of trions in
external magnetic fields requires fully symmetry-consistent,
Landau-level–complete frameworks capable of handling strong
Landau-level mixing and the nonseparability of center-of-mass and
internal motion. Third, anisotropic and quasi-1D
monolayers, including phosphorene and transition-metal
trichalcogenides, offer promising platforms for exploring
direction-dependent few-body correlations and dimensional crossover
effects.

Equally important is the experimental frontier. High-quality,
charge-tunable, and encapsulated devices are essential for resolving
trion fine structure, intervalley configurations, and magnetotrion
states in emerging material families such as Xenes and TMTCs.
The continued interplay between advanced spectroscopy and rigorous
few-body theory will be crucial for refining interaction models beyond
simplified screening approximations.

In a broader perspective, trions provide a paradigmatic realization of
few-body Coulomb physics embedded within a many-body solid-state
environment. Their study not only deepens our understanding of
correlated quasiparticles in low dimensions but also establishes a
bridge between nuclear-style few-body methods and condensed-matter
phenomena. As materials platforms expand and computational techniques
continue to mature, trion physics is poised to remain a central arena
for testing and advancing few-body concepts in quantum materials.

\section*{Acknowledgments}

R.~Ya.~K. is grateful to K.~Ziegler, A.~Spiridonova, and Sh.~M.~Tsiklauri for valuable discussions.

\end{document}